\documentclass[fleqn,usenatbib]{mnras}

\usepackage{newtxtext,newtxmath}
\usepackage[T1]{fontenc}

\DeclareRobustCommand{\VAN}[3]{#2}
\let\VANthebibliography\thebibliography
\def\thebibliography{\DeclareRobustCommand{\VAN}[3]{##3}\VANthebibliography}

\usepackage{graphicx}	% Including figure files
\usepackage{amsmath}	% Advanced maths commands
\usepackage{float}
\usepackage{graphicx}
\usepackage{dsfont}
\usepackage{amsmath}

\usepackage{xcolor}
\usepackage{xparse}
\usepackage{listings}

\usepackage{graphicx} % Required for inserting images
\usepackage{caption}
\usepackage{subcaption}
\usepackage{tabularray}
\usepackage[labelfont=bf]{caption}
\usepackage{gensymb}
\usepackage{setspace}

\title[HCFs in Elliptical Galaxies]{Hidden Cooling Flows in Elliptical Galaxies}

\author[L. R. Ivey et al.]{
L. R. Ivey,$^{1}$\thanks{E-mail: li247@cam.ac.uk}
A. C. Fabian,$^{1}$
J. S. Sanders,$^{2}$
C. Pinto,$^{3}$
G. J. Ferland,$^{4}$
S. Walker,$^{5}$
J. Jiang$^{6, 1}$
\\
$^{1}$ Institute of Astronomy, University of Cambridge, Madingley Road, Cambridge CB3 0HA, UK
\\
$^{2}$ Max-Planck-Institut fur extraterrestrische Physik, Giessenbachstrasse 1, 85748 Garching, Germany\\
$^{3}$ INAF-IASF Palermo, Via U. La Malfa 153, I-90146 Palermo, Italy\\
$^{4}$ Department of Physics and Astronomy, The University of Kentucky, Lexington, KY 40506, USA\\
$^{5}$ Department of Physics and Astronomy, University of Alabama in Huntsville, Huntsville, AL 35899, USA\\
$^{6}$ Department of Physics, University of Warwick, Gibbet Hill Road, Coventry CV4 7AL, UK\\
}

\date{Accepted XXX. Received YYY; in original form ZZZ}

\pubyear{\the\year{}}

\begin{document}
\label{firstpage}
\pagerange{\pageref{firstpage}--\pageref{lastpage}}
\maketitle

\begin{abstract}
The radiative cooling time of hot gas in the cool cores of many galaxy clusters and massive elliptical galaxies drops in the centre to below $10^8$ years. The mass cooling rates inferred from simple modelling of X-ray observations of these objects are very low, indicating that either AGN feedback is tightly balanced or that soft X-rays from cooling gas are somehow hidden from view. An intrinsic absorption model developed for application to galaxy clusters is used here to search for hidden cooling flows (HCFs) in seven nearby elliptical galaxies. Mass cooling rates of $\sim$0.5--8 \ $\text{M}_{\odot}$ yr\textsuperscript{-1} are found in each galaxy. The absorbed cooling flow luminosity is in agreement with the observed Far Infrared (FIR) luminosity in each case, indicating absorbed emission is energetically capable of emerging in the FIR band. An observed lack of agreement between HCF rates and normal star formation rates suggests the cooled material must have an alternative fate, with low-mass star formation considered as the primary outcome.
\end{abstract}


%%%%%%%%%%%%%%%%%%%%%%%%%%%%%%%%%%%%%%%%%%%%%%%%%%

%%%%%%%%%%%%%%%%% BODY OF PAPER %%%%%%%%%%%%%%%%%%

\section{Introduction}\label{Introduction}

The gas clouds from which galaxies and clusters form are heated by energy released during their gravitational collapse. The hot atmosphere loses energy by emission of radiation, primarily in X-rays from bremsstrahlung. For X-ray radiation, the radiative cooling time $t_\text{cool} \propto T^{1/2}/n$ falls with increasing gas density $n$ and decreasing temperature $T$ \citep{Fabian1994}. Observations of the sharply-peaked X-ray surface brightness distributions of clusters indicate that gas density increases towards the centre of the cluster, resulting in a correspondingly short cooling time, much less than the age of the cluster (approximated by $t_\text{a} \sim H_0^{-1}$); objects with this property are said to have \textit{cool cores} \citep{Sanders2023}. 

As gas in the centre of the cluster cools, the pressure of the overlying gas causes more gas to flow inwards \citep{Fabian1994}. This is referred to as a \textit{cooling flow} \citep{CowieBinney1977, fabiannulsen1977}, and X-ray observations have indicated such flows are common in the centres of galaxy clusters and massive elliptical galaxies. 
 
The rate at which gas cools out of the X-ray band is related to the mass deposition rate $\dot{M}$, which can be estimated from the emitted X-ray luminosity $L$ using
\begin{equation}
    \centering
    \dot{M} = \frac{2}{5} \frac{L\mu m_{\text{H}}}{kT},
    \label{Equation 1}
\end{equation}
where $T$ is the temperature and $\mu m_{\text{H}}$ is the mean mass per particle \citep{HCF1, Sanders2023}.
X-ray observations of cooling flows indicate they should contain a range of gas temperatures, from an ambient cluster temperature down to temperatures below $10^6$ K at which gas does not emit X-rays.

However, in many galaxy clusters, there is little direct observational evidence for cool X-ray emitting gas at temperatures of 1 keV and below. Observed star formation rates are orders of magnitude smaller than the naive mass cooling rate predicted from \hyperref[Equation 1]{Equation 1} assuming no intrinsic absorption of emitted X-rays or obscuration by cool plasma \citep{1987JFN, McDonald2018}. Furthermore, XMM-Newton RGS spectrometer data have demonstrated that a standard cooling flow model overpredicts the Fe XVII emission lines expected from the lowest temperatures in X-rays \citep{Peterson2003}. Not only is there a failure to observe the cooling gas, but the fate of any cooled material was also unclear; this is referred to as the \textit{Cooling Flow Problem} (see \cite{McDonald2018} for details).

Early-type elliptical galaxies bear many similarities to the cool cores of galaxy clusters. \cite{Lakhchaura18} and \cite{Babyk2019} discuss the temperature and density profiles in elliptical galaxies as well as cooling timescales, noting their similarity to brightest cluster galaxies (BCGs). Towards the centre of an elliptical galaxy, the mean temperature profile approaches isothermal, the gas is coolest and densest, and the radiative cooling time is shortest; these features are indicative of a cooling flow. We would therefore expect to see gas rapidly cooling below 1 keV. However, while Fe XVII and O VII lines indicative of low-temperature gas are often present in the X-ray spectra of elliptical galaxies \citep{SandersFabian2011, Pinto_2014, PintoEtAl2016}, they are significantly suppressed compared to what we would expect to observe without absorption \citep{Sanders2023}. Consequently, very little gas is directly observed at temperatures below 1 keV compared to predictions, demonstrating that the Cooling Flow Problem is operating in elliptical galaxies. 

To address this issue, it is necessary to consider what is happening to the missing soft X-ray emission (0.1--1 keV). There are several possible explanations. For example, the cooling may be suppressed by heating mechanisms, with AGN feedback as the most likely candidate \citep{2005Bruggen}. However, the presence of some iron lines in the spectra of galaxy clusters and elliptical galaxies indicate mild cooling flows at least are present \citep{PintoEtAl2016}, so AGN feedback cannot completely quench radiative cooling in these objects. Furthermore, heating by the AGN would need to be tightly balanced with the cooling flows to ensure that gas is not under- or over-heated compared to observations, presenting a significant fine-tuning problem. 

While there is unambiguous evidence of AGN feedback, photoelectric absorption \citep{WhiteEtAl1991, JohnstoneEtAl1992, Fabian1994abs} is an another process which must also be considered in resolving this problem. In order for most of the soft X-rays from cooling gas to be absorbed, the cooling gas must be closely interleaved with cold gas. The intrinsic multilayer absorption model (IAM) based on this idea, introduced by \cite{Allen&Fabian1997spatial}, was found to fit well to both low-resolution \citep{Allen&Fabian1997spatial, Allen2000} and high-resolution X-ray spectra \citep{HCF1} of the cool cores.

\begin{figure}
     \centering
     \includegraphics[width=0.48\textwidth]{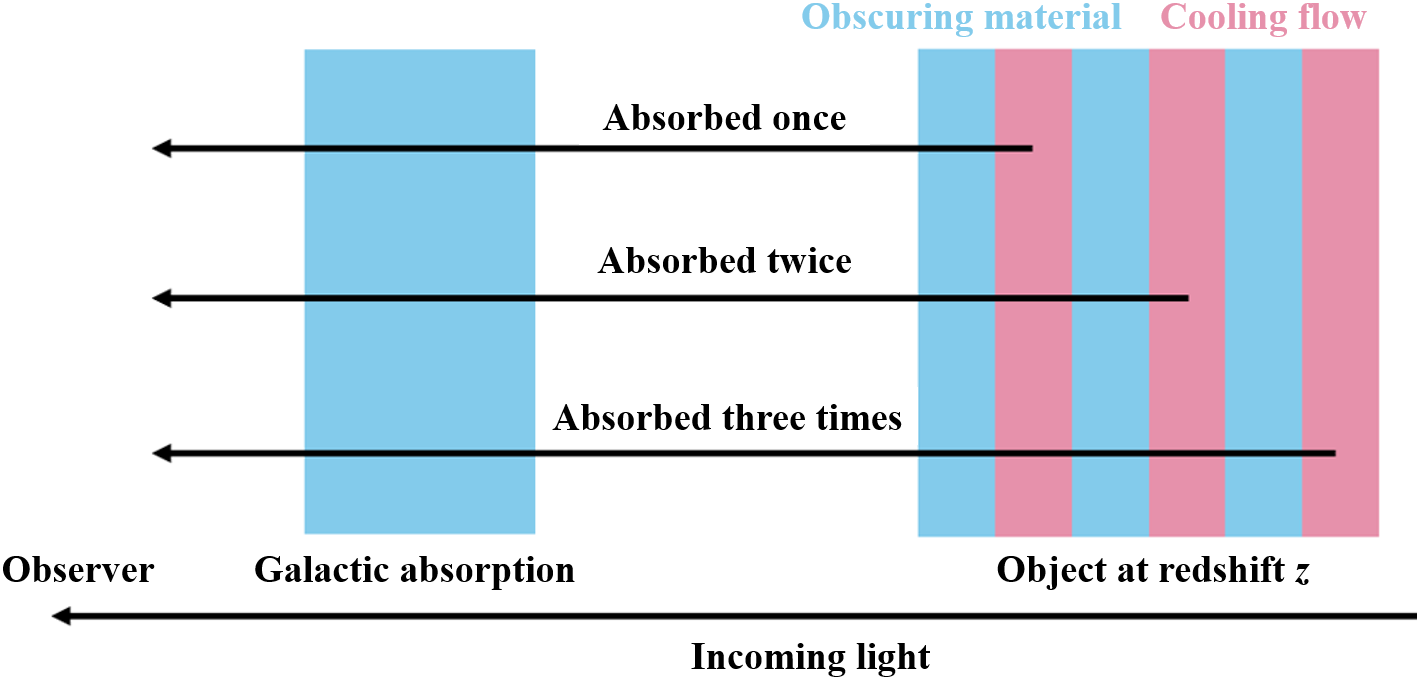}
        \caption{A schematic diagram of a multilayer cooling flow, with three absorbing sheets and three cooling flow sheets. In the object at redshift $z$, blue columns represent sheets of absorbing gas, and pink columns represent sheets of radiatively cooling gas. There is additional Galactic absorption along the line of sight between the observer and the object. Black arrows originating from the object represent emission from the associated sheet of cooling flow. This Figure was adapted from Figure 5 in \citealt{Liu2021}.}
    \label{iam}
\end{figure}

\hyperref[iam]{Figure 1} illustrates the concept of multilayer absorption, in a simple case with only three layers each of emission and absorption. In reality, it is necessary to consider significantly more layers than this. Where cold absorbing gas is interspersed with soft X-ray emitting gas, some emission will be viewed unabsorbed; however, the majority of the emission is observed through considerable absorption. To avoid introducing too much complexity, the possibility that the ratio of emission to absorption could be some function of depth or emission temperature is neglected here, in favour of making a simple model to obtain limits on mass cooling rates.

Furthermore, it is proposed that the absorbing layers emit the energy absorbed from soft X-rays at longer wavelengths. This means that the energy from the `missing' soft X-ray radiation could instead be emerging at significantly longer wavelengths, in the UV to far-infrared (FIR) bands. The FIR luminosity of a galaxy or cluster can thus be used to constrain the mass cooling rates estimated from the IAM.

Along with this qualitative depiction, a mathematical prescription of the model has been developed \citep{1992CrawfordFabian, Allen&Fabian1997spatial}.  Assume that each layer of emission at wavelength $\lambda$ contributes a flux $\Delta F$. Additionally, each layer of absorption of column density $\Delta N$ transmits a fraction $f = e^{-\sigma \Delta N}$ of the flux, where $\sigma$ is the appropriate photoelectric absorption cross section at that wavelength. When the total emitted flux is $F_e = n\Delta F$, the total flux observed from $n$ layers, $F_t$, is 
\begin{equation}F_t = \Delta F + f \Delta F + f^2 \Delta F + ... + f^n \Delta F.
\label{Equation2}
\end{equation}
The above is a geometric series, which reduces in the limit of large $n=N_\text{H}/\Delta N$ to give a total observed flux 
\begin{equation}F_t = \frac{F_e \left(1-e^{-\sigma N_{\textsc{H}}}\right)}{\sigma N_{\textsc{H}}},\label{Equation3}\end{equation}
where $N_\text{H}$ is the total column density of intrinsic absorption. 

There is a further component of absorption which arises due to Galactic absorption along the line of sight between the observer and the object, with associated column density $N_\text{H}'$; this is applied on top of the intrinsic multilayer component just discussed.  In an X-ray spectral fitting program such as XSPEC, the complete model is implemented as a multiplicative model. As the hidden cooling flow (HCF) rate $\dot{M}$ is a parameter of the model, fitting the model to the spectrum of an object allows this rate to be calculated for the object in question.

The intrinsic multilayer absorption model has been successful in demonstrating HCFs in a diverse range of sources. Work on ASCA and ROSAT spectra found that not only does the IAM fit well, it can also allow significant cooling flows \citep{Allen2000, AllenEtAL2001}. More recently, the IAM has been revived and   successfully applied to the  massive cluster RXJC1504.1-0248 \citep{Liu2021}. HCFs were found to be common in the cool cores of elliptical galaxies, groups and clusters \citep{HCF1, HCF2, HCF3}. In particular, galaxy clusters were studied, finding mass cooling rates of hundreds of solar masses per year, centred on the brightest cluster galaxy (BCG). 

While previous work on the Cooling Flow Problem has mainly focused on galaxy clusters, individual elliptical galaxies have also been investigated \citep{1986Thomas, 1987SarazinWhite}. Even before the IAM was first applied, \cite{Rangarajan1995} had identified an absorbed cooling flow in NGC 1399, the central galaxy of the nearby Fornax cluster. Recently, in \cite{HCF2}, HCFs of $\sim$1--2$\ \text{M}_\odot$ yr\textsuperscript{-1} were found by fitting the IAM to spectra of the nearby elliptical galaxies M49 and M84. \cite{HCF3} took this further, investigating a few more elliptical galaxies, finding HCFs of $\sim$1--10$\ \text{M}_\odot$ yr\textsuperscript{-1} in each; this scales appropriately to the cluster HCFs when galaxy mass is considered relative to cluster mass. These findings further motivate the extension of the search for hidden cooling flows to within individual galaxies, with the aid of high-resolution X-ray spectra provided by the XMM-Newton RGS.

Finding these HCFs then raises a further question: what is happening to the cooled gas from an HCF? The two main outcomes to consider are (i) the redistribution and reheating of cooled gas by AGN feedback, and (ii) the fragmentation and collapse of clouds of cooled gas, forming low-mass stars and brown dwarfs. These possibilities are not mutually exclusive, but the fine tuning problem associated with AGN feedback has already been pointed out. Consequently, low-mass star formation is currently the leading explanation for the ultimate fate of cooled gas from HCFs \citep{Fabian2024}.

In this work, we implement the intrinsic multilayer absorption model to search for significant HCFs with gas cooling down to temperatures of 0.1 keV in a sample of 7 nearby elliptical galaxies: NGC 1316 (\textit{Fornax A}), NGC 1332, NGC 1404, NGC 4552 (\textit{M89}), NGC 4636, NGC 4649 (\textit{M60}) and IC 1459. These galaxies were chosen based on availability of RGS spectral data and the proximity of the galaxies themselves. Deprojection of Chandra imaging data are shown for 5 of our objects in  \cite{Lakhchaura18}. All  5 have central cooling times below $10^8$ years; these are upper limits given that the central temperature in that work is typically $\sim$ 0.9 keV. 

A key assumption in our work is that once gas cools below 0.1 keV, it will continue to cool to $10^4$ K and below. Hidden cooling flow rates are calculated for each galaxy using spectral fitting and the potential fate of the cooling gas is assessed. 

To make a full comparison of the rates of all heating and cooling processes in these objects, it would be necessary to also account for processes such as stellar mass loss and supernovae heating, as was done in early work by \cite{1986Thomas} and \cite{1987SarazinWhite}. We will revisit the contribution from stellar mass loss in \hyperref[Further]{Section 6}. However, this work primarily considers and identifies radiative cooling components obtained from modelling RGS spectra, which lack the spatial resolution necessary for a more detailed investigation.

Throughout this work, the following cosmological parameters are assumed, unless stated otherwise: $H_0 = 70$ km s\textsuperscript{-1} Mpc\textsuperscript{-1}, $q_0 = 0$, $\Lambda_{\text{0}} = 0.73$, and $\Lambda_{\text{m}} = 0.27$.

\section{Spectral Analysis}\label{SA}

This study uses XMM-RGS spectra of 7 nearby elliptical galaxies. The \textit{XMM–Newton} observatory, which produced these spectra, is composed of X-ray imaging cameras and gratings. The RGS is a pair of dispersive spectrometers \citep{denHerder2001} which provide the spectrum of a slice across the centre of a nearby elliptical galaxy or galaxy cluster. The observation IDs for each source listed in \hyperref[Table1]{Table 1} were processed initially with the \textsc{rgsproc} RGS pipeline from the XMM Science Analysis Software \citep{Gabriel2004}, using the source positions listed. We extracted events from 95 per cent of the PSF, corresponding to an extraction width of approximately 1.7 arcmin, and 95 per cent of the pulse-height distribution. The count rate, in bins of 200 seconds, was measured for each RGS on CCD 9, with flag values of 8 or 16, with an absolute cross dispersion angle of less than $1.5 \times 10^{-4}$. Time periods in the dataset where this exceeded $0.3 \ \mathrm{s}^{-1}$ were excluded from the analysis. Background spectral based on models was created using \textsc{rgsbkgmodel}. RGS 1 and 2 each have an electronically dead chip covering 10.4--13.8 \AA \ and 20--24 \AA \ respectively, so the first order spectra of the two RGS instruments were combined using \textsc{rgscombine}, ensuring all wavelengths were covered. The spectra were grouped to have a minimum of 20 counts per spectral channel.

{\renewcommand{\arraystretch}{1.4}
\begin{table*}
\normalsize
\centering
\begin{tblr}{c c c c c}
 \hline
 Target   & Exposure & RA & Dec & Observation ID(s) \\
 \hline
 & ks & $\degree$ & $\degree$ & \\
 \hline
NGC 1316 & $333.2$ & $ 50.6738$ & $-37.2082$ & 0302780101 0502070201 \\ NGC 1332 & $127.1$ & 51.5719 & $-21.3352$ & 0304190101 \\ NGC 1404 & $343.3$ & $ 54.7163$ & $-35.5944$ & 0304940101 0781350101 \\ NGC 4552 & $ 56.4$ & $188.9159$ & $+12.5563$ & 0141570101 \\ NGC 4636 & $122.4$ & $190.7076$ & $ +2.6878$ & 0111190101 0111190701 \\ NGC 4649 & $377.5$ & $190.9166$ & $+11.5527$ & 0021540201 0021540401
0502160101 0845140201 \\
IC 1459  & $317.1$ & $344.2942$ & $-36.4622$ & 0135980201 0760870101 \\
\hline
\end{tblr}
\caption{The listed exposure times are the cleaned exposure times for the first order spectra, summing the RGS1 and 2 exposure times. The positions give the source position given to \textsc{rgsproc}.}
\label{Table1}
\end{table*}
}

Initially, the spectra were investigated over the wavelength range 8--22 \AA, which is where background is minimised. However, in several galaxies in our sample, notable O VII lines were observed at $\sim$22 \AA. We extended the wavelength range studied to include up to 25 \AA \ so that the O VII lines could clearly be seen and fitted to the model.

In the X-ray spectral-fitting program XSPEC \citep{Arnaud1996}, the intrinsic multilayer absorption model described in \hyperref[Introduction]{Section 1} is defined as
\textsc{tbabs(gsmooth*apec+gsmooth(partcov*mlayerz)mkcflow)}.
This has two terms: a single-temperature thermal gas, and an intrinsically absorbed cooling flow.

\textsc{tbabs} is the Tuebingen-Boulder ISM model, which calculates the cross-section for X-ray absorption by the ISM \citep{tbabsref}; this component describes the Galactic absorption in the direction of the target and it applies to both terms. The two emissive components of the model (\textsc{apec} and \textsc{mlayerz}) each have a separate Gaussian smoothing applied to them, described by \textsc{gsmooth}. This compensates for the spatial extent of the source and intrinsic turbulent blurring; cooler gas is typically seen to show narrower emission lines, in agreement with the location being confined to a smaller central region \citep{PintoEtAl2016}. \textsc{apec} is the constant temperature thermal emission model, which represents the outer galactic gas. \textsc{partcov} describes the percentage of the galaxy which is covered by absorption: the covering fraction is 1 when all of the cooling flow component is absorbed, and it is zero when none is absorbed. This model does not assume any particular geometry for the absorbed and unabsorbed components, beyond the idea that the unabsorbed gas lies outside of the absorbed component. \textsc{mlayerz} is the intrinsic absorption model described in \hyperref[Introduction]{Section 1}; here it is implemented as a multiplicative model using \textsc{mdefine}\footnote{mdefine mlayer (1-ztbabs(nh,z))/(-ln(ztbabs(nh, z))) :mul}. Finally, \textsc{mkcflow} is a cooling flow model, whose hotter temperature is given by the \textsc{apec} component, and it describes cooling from that temperature down to 0.1 keV. This spectral model allows the direct measurement of total mass cooling rate of both unabsorbed and absorbed components. 

Details of the target galaxies are shown in \hyperref[Table2]{Table 2}. Each galaxy has a fixed redshift and Galactic absorption. The Galactic absorption in each case was calculated using the `nh' FTOOL in XSPEC, which returns Galactic hydrogen column density for an input RA and Dec; it is based upon HI maps by \cite{Dickey1990} and \cite{Kalberla2005}. The Gaussian smoothing is also often fixed for spectral fitting, as sensitivity to this parameter is low.

{
\begin{table}
\footnotesize
\centering
 \begin{tabular}{c c c c c} 
 Target & RA & Dec & Redshift & Mean Distance\\ 
 \hline
 & $\degree$ & $\degree$ & $10^{-3}$ & Mpc\\
 \hline
 NGC 1316 & 50.67 & -37.21 & 6.01 & 19.2 $\pm$ 0.6\\
 NGC 1332 & 51.57 & -21.34 & 5.40 & 22.8 $\pm$ 3.2\\
 NGC 1404 & 54.72 & -35.59 & 6.49 & 19.3 $\pm$ 0.4\\
 NGC 4552 & 188.92 & 12.56 & 1.13 & 16.5 $\pm$ 0.5\\
 NGC 4636 & 190.71 & 2.69 & 3.13 & 16.3 $\pm$ 0.6\\
 NGC 4649 & 190.92 & 11.55 & 3.70 & 16.7 $\pm$ 0.4\\
 IC 1459 & 344.29 & -36.46 & 6.01 & 26.2 $\pm$ 2.1\\
 \hline
 
 \end{tabular}
 \caption{Data for the objects used in this study. Mean distance measurements specified are redshift-independent. All data courtesy of the NASA Extragalactic Database. }
 \label{Table2}
\end{table}
}

\begin{figure}
\centering
\includegraphics[width=0.48\textwidth]{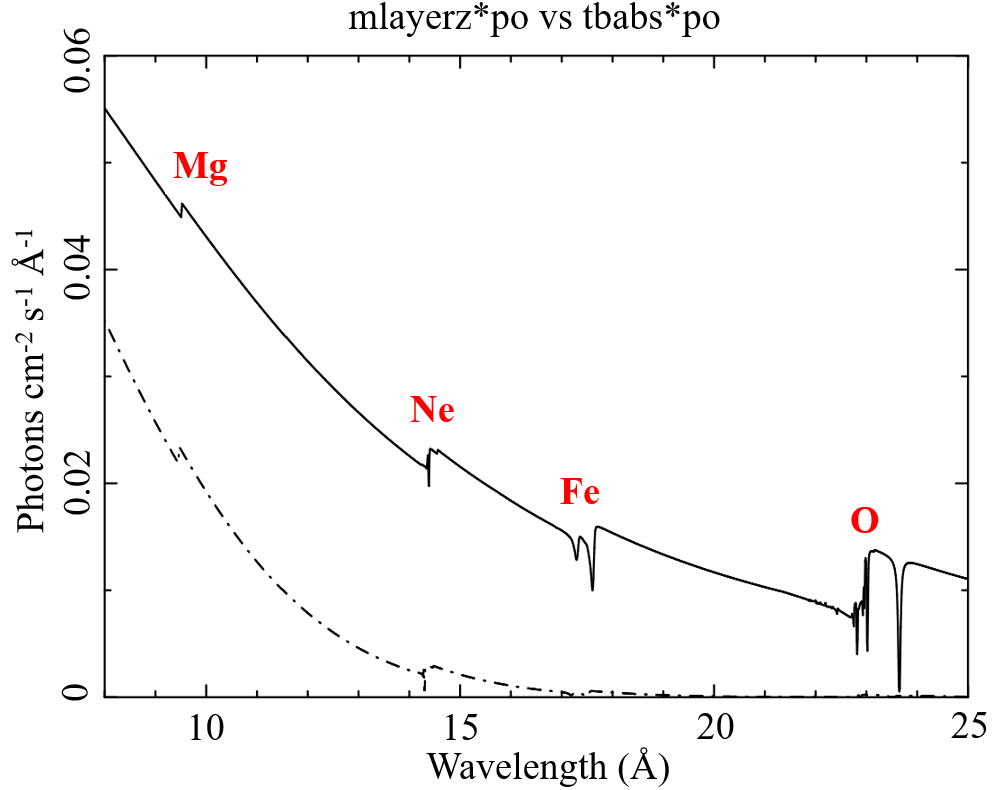}
    \caption{Plot comparing two simple models (\textsc{mlayerz*powerlaw}, the upper solid line, and \textsc{tbabs*powerlaw}, the lower dashed line), each with intrinsic absorption column density $10^{22}$ cm\textsuperscript{-2} applied to a power-law index of 2. This demonstrates the fraction of transmitted emission at each wavelength and the absorption edges associated with individual elements. Note that longer wavelengths are subject to greater absorption, and the O absorption edge is at $\sim$23 \AA.}
    \label{Figure2}
\end{figure}

Once these parameters are fixed, the model is fitted in XSPEC using $\chi^2$ statistics. In fitting the model to galaxy spectra, there are three crucial parameters: $N_{\text{H}}$, $CFrac$ and $\dot{M}$. 

\begin{figure}
     \centering
     \includegraphics[width=0.48\textwidth]{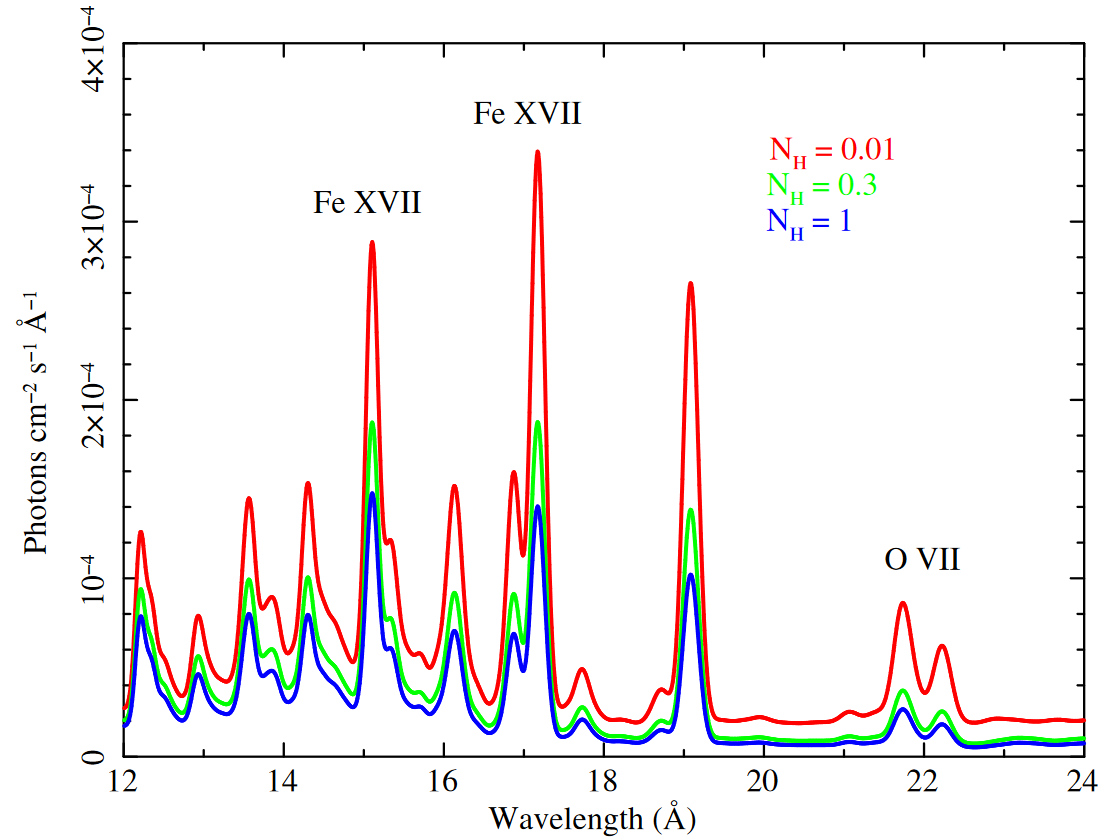}
        \caption{Intrinsic multi-layer absorption models fitted for a range of $N_\text{H}$ values, specified in units of $10^{22}$ cm\textsuperscript{-2}. All other model parameters were fixed. Note the suppression of the Fe XVII lines at 15 and 17 \AA, as well as the O VII lines at 22 \AA. The effect of line suppression is more dramatic at longer wavelengths. For a more complete identification of the spectral lines, see Figure 3 in \citealt{SandersFabian2011}.}
    \label{Figure3}
\end{figure}

$N_{\text{H}}$ is the total column density of intrinsic absorption. \hyperref[Figure2]{Figure 2} demonstrates how the intrinsic multilayer absorption model \textsc{mlayerz} affects the fraction of emission transmitted at each wavelength. Longer wavelengths are subject to increased absorption, up to the absorption edge at  $\sim$23\ \AA. \hyperref[Figure3]{Figure 3} illustrates the effect of increasing $N_\text{H}$ on a model galaxy spectrum. Lower energy components experience more absorption, so a greater value of $N_{\text{H}}$ acts to suppress emission at all wavelengths, but particularly at wavelengths in the range 15--23 \AA. In \hyperref[Figure3]{Figure 3}, three spectral lines in particular are labelled: the resonance and forbidden lines of Fe XVII at $\sim$15 \AA \ and $\sim$17 \AA \ respectively, and the O VII triplet at $\sim$22 \AA \ (in particular, the resonance, intercom and forbidden lines at 21.6, 21.8 and 22.1 \AA \ respectively). These lines can be particularly prominent as the ionisation states they correspond to have noble gas-like electron configurations (neon-like in the case of Fe XVII; helium-like in the case of O VII) and hence are longer-lived in a cooling flow than other, less stable ionisation states. Fe XVII lines are often prominent in the spectra of elliptical galaxies, indicative of gas at temperatures of 0.5--0.75 keV. However, the O VII lines, which are indicative of gas at temperatures of 0.25 keV and below, are often weak if they are present at all in the spectra. The transmission profile of intrinsic absorption shown in \hyperref[Figure2]{Figure 2} gives a mechanism to explain these observations, as it demonstrates why O VII lines are more suppressed by absorption than the Fe XVII lines. It should also be noted that the O VII lines lie close to the oxygen absorption edge, and the sensitivity is reduced by the dead chip in the 20--24 \AA \ band of RGS2.

$CFrac$, the covering fraction, describes the proportion of emission which is covered by intrinsic absorption. An increase in $CFrac$ acts to suppress emission across \textit{all} wavelengths, as a greater area is subject to the same total column density of intrinsic absorption. This distinction between $N_\text{H}$ and $CFrac$ breaks their degeneracy, so to achieve the best possible fit it is necessary to optimise over both of these parameters. For the objects studied in this paper, we find that CFrac is close to unity (see the results presented in \hyperref[Table4]{Table 4}).

$\dot{M}$, the hidden mass cooling rate, is the third key parameter; it is strongly driven by the relative strength of emission lines. Optimisation over $N_{\text{H}}$ and $CFrac$ enables an estimate of $\dot{M}$ to be determined.

The resulting spectra are shown in Figures 4--10 in the \hyperref[Results]{next section}, together with contour plots of absorbed mass cooling rate versus intrinsic column density and covering fraction obtained using the \textit{steppar} routine in \textsc{xspec}. The best-fit model parameters are summarised in \hyperref[Table4]{Table 4}, also in the next section.

In addition to the IAM, two additional models were also fit to the data: a Single-Temperature model and a Two-Temperature model. These models are described in \hyperref[AppendixA]{Appendix A} and \hyperref[Table3]{Table 3}. This was done to demonstrate the goodness of fit of the IAM compared to simpler models which do not contain cooling flows, and to give confidence in the calculated hidden mass cooling flow rates. A further check was performed by refitting the models using \textit{cstat} as the fit statistic; see \hyperref[AppendixB]{Appendix B} for details.

A further model adjustment was necessary to account for nuclear emission in the case of IC 1459 (see \hyperref[Section3.7]{Section 3.7}).

\section{Results}\label{Results}

\begin{table*}
\normalsize
\centering
 \begin{tabular}{c c c c c} 
 \hline
 Target & \multicolumn{4}{c}{$\chi^2$/dof} \\
 \hline
  & Single-Temperature & Two-Temperature & Unabsorbed Cooling Flow & Intrinsic Absorption \\
 \hline
 NGC 1316 & 1406/488 & 615/484 & 574/484 & 495/484\\
 NGC 1332 & 200/124 & 129/120 & 170/122 & 135/121\\
 NGC 1404 & 1302/803 & 1164/801 & 1266/801 & 1173/800\\
 NGC 4552 & 197/108 & 156/106 & 169/106 & 134/105\\
 NGC 4636 & 988/612 & 785/610 & 924/610 & 895/609\\
 NGC 4649 & 1111/878 & 1075/876 & 1113/876 & 1062/875\\
 IC 1459 & 499/315 & 408/312 & 315/312 & 311/311\\
 \hline
 
 \end{tabular}
 \caption{Comparison of $\chi^2$/dof values for different model fits to each galaxy. For a description of the Single-Temperature and Two-Temperature models, see \hyperref[AppendixA]{Appendix A}. While the Two-Temperature model is occasionally a slightly better fit than the Unabsorbed Cooling Flow model, the Intrinsic Absorption model is always statistically preferred overall with respect to the unabsorbed model.}
 \label{Table3}
\end{table*}

{\renewcommand{\arraystretch}{1.4}
\begin{table*}
\normalsize
\centering
 \begin{tblr}{c c c c c c c c c c}
 \hline
 Target & $N_\text{H}'$ & $kT$ & $Z$ & $Norm$ & $CFrac$ & $N_\text{H}$& $\dot{M}$ & $\dot{M}_\text{u}$\\ 
 \hline
  & $10^{22}$ cm\textsuperscript{-2}& keV & $Z_{\odot}$& $10^{-4}$ &  & $10^{22}$ cm\textsuperscript{-2}& $M_\odot$ yr\textsuperscript{-1} &  $M_\odot$ yr\textsuperscript{-1}\\
 \hline
 NGC 1316 & 0.0199 & $0.98_{-0.10}^{+0.07}$ & $0.22\pm0.04$ & $0.47$ & $0.91_{-0.13}^{+0.04}$ & 3.6 & $4.2_{-2.5}^{+3.1}$ & $0.49^{+0.01}_{-0.04}$\\
 NGC 1332 & 0.0214 & $0.68_{-0.05}^{+0.04}$ & $0.11_{-0.03}^{+0.04}$ & 2.43 & $1_{-0.03}$ & 4.0 & $5.8^{+3.6}_{-3.4}$ & $0.23\pm0.05$\\
 NGC 1404 & 0.0138 & $0.69\pm0.02$ & $0.21\pm0.01$ & 19 & $0.97\pm0.03$ & 2.3 & $7.7^{+0.9}_{-1.0}$ & $0.55^{+0.11}_{-0.12}$\\
 NGC 4552 & 0.0266 & $0.70_{-0.10}^{+0.03}$ & $0.07^{+0.02}_{-0.01}$ & 6.6 & $0.97^{+0.16}_{-0.03}$ & 4.4 & $0.65\pm0.12$ & $0.034^{+0.004}_{-0.005}$\\
 NGC 4636 & 0.0182 & $0.75\pm0.01$ & $0.27\pm0.02$ & 31 & $1_{-0.18}$ & 0.4 & $0.89_{-0.11}^{+0.25}$ & $0.28\pm0.05$\\
 NGC 4649 & 0.0202 & $0.90\pm0.01$ & $0.33\pm0.02$ & 22 & $1_{-0.01}$ & $10.06$ & $3.4_{-2.4}^{+1.3}$ & negligible\\
 IC 1459 & 0.00967 & $0.73_{-0.08}^{+0.16}$ &$0.11_{-0.02}^{+0.04}$ & 1.3 & $0.91_{-0.51}^{+0.09}$ & 6.5 & $2.7_{-2.3}^{+3.3}$ & $0.30_{-0.05}^{+0.06}$\\
 \hline
\end{tblr}
\caption{Model parameters for the analysis. The redshifts for each galaxy are as given in \hyperref[Table2]{Table 2}. $kT$ is the temperature of the \textsc{apec} model component; $Z$ is the abundance relative to solar abundance. $N_\text{H}$ is the column density of intrinsic absorption, while $H_\text{H}'$ is the column density associated with absorption in our Galaxy. Where $Cfrac$ was fitted as 1, there is no upper error, so only lower bounds on error are given. The errors represent 90\% confidence intervals using $\chi^2$.}
\label{Table4}
\end{table*}
}
{\renewcommand{\arraystretch}{1.2}}

All galaxies studied here are best fit with a Hidden Cooling Flow. All require a covering fraction $> 0.9$, and all but 2 allow for a mass cooling rate of between 2 and $8 \ \text{M}_\odot$ yr\textsuperscript{-1}.
Each galaxy can allow for a hidden cooling flow of several solar masses per year. Of the 7 sources studied, all but two require a best fit covering fraction of 0.95 or more, which emphasizes that the emission is in fact `hidden'. The intrinsic column densities range from $0.4 \times 10^{22}$ to $10.2 \times 10^{22}$ cm\textsuperscript{-2}. The full results of spectral fitting are summarised in \hyperref[Table4]{Table 4}.

The spectra were also re-fitted with $CFrac$ set to zero (this model will be referred to as the \textit{Unabsorbed Cooling Flow Model}) to determine the unabsorbed mass cooling rate, $\dot{M}_\text{u}$, which is the mass cooling rate allowed by the observed spectrum assuming there is no absorption covering the galaxy (see the last column of \hyperref[Table4]{Table 4}). As expected, the unabsorbed mass cooling rate is much smaller than than the absorbed mass cooling rate $\dot{M}$, generally by one order of magnitude. These findings are summarised in the context of individual galaxy properties in Sections 3.1--3.7.

A full comparison of $\chi^2$/dof for the models studied is presented in \hyperref[Table3]{Table 3}. Across all of the galaxies, the IAM consistently has a lower $\chi^2$ than any of the other models, indicating it is statistically preferred over a Single-Temperature, Two-Temperature or Unabsorbed Cooling Flow model.

For completeness, we also tested using \textit{cstat} as a fit statistic, and no significant change in fit parameters was found. For a summary, see \hyperref[TableB]{Table B1}.

\subsection{NGC 1316}

NGC 1316, also known as \textit{Fornax A}, is a radio galaxy located at $z = 0.00601$, near the edge of the Fornax cluster \citep{Ferguson1989}. At 1400 MHz it is one of the brightest radio sources in the sky, with an active galactic nucleus and two radio lobes; it also has prominent dust lanes (see \hyperref[11a]{Figure 11(a)} and \citealt{1980}). It is suspected to have built up through the merger of several smaller galaxies, which could have fuelled the growth of its central supermassive black hole, of mass $1.5 \times 10^8$ \(\textup{M}_\odot\) \citep{Nowak_2008}. This SMBH powers a low-luminosity AGN, with an X-ray luminosity $L_\text{X}=5 \times 10^{39}$ erg s\textsuperscript{-1} (0.3--8 keV). Like all the galaxies studied here, NGC 1316 also contains a large number of X-ray bright point sources, which are not bright enough for individual spectral analysis \citep{KimFabbiano2003}, and likely have featureless spectra (see \hyperref[Correction]{Section 5}).

\begin{figure}
\centering
\begin{subfigure}{0.47\textwidth}
\includegraphics[width=\textwidth]{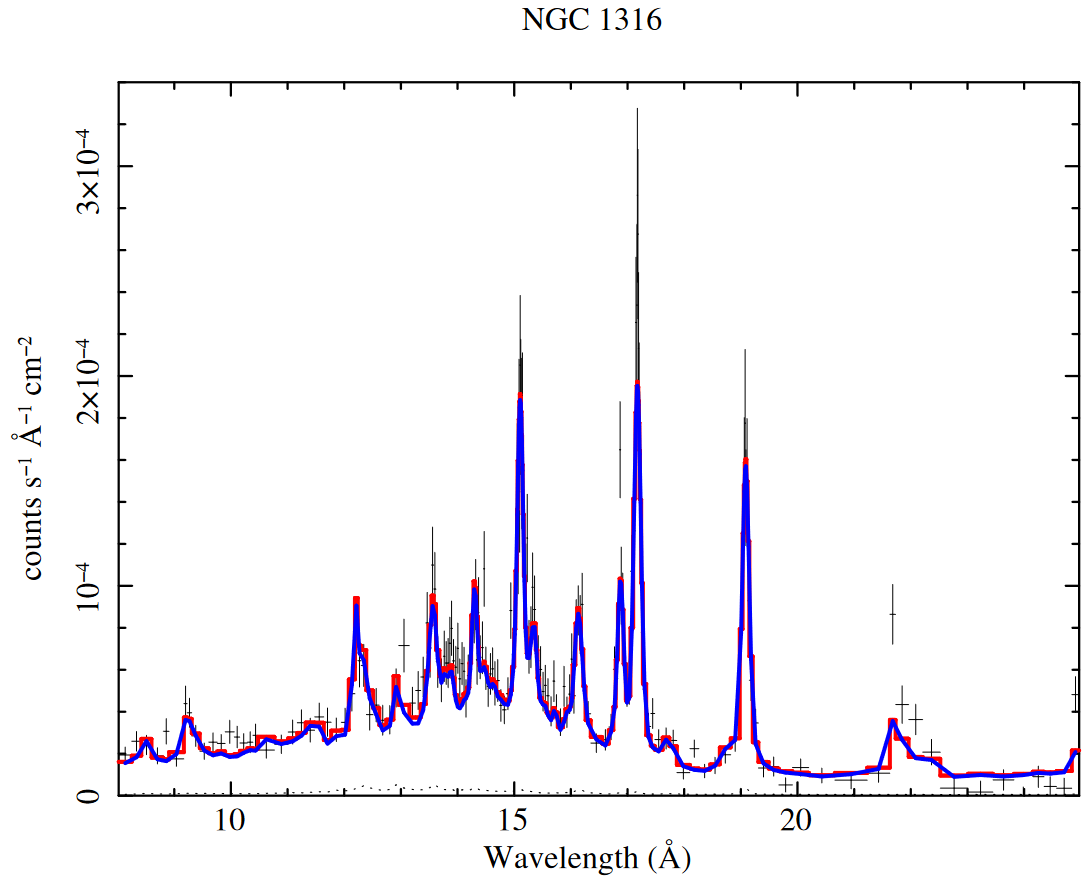}%
\caption{Spectrum}
\label{4a}
\end{subfigure}
 \begin{subfigure}{.47\textwidth}
\includegraphics[width=1\textwidth]{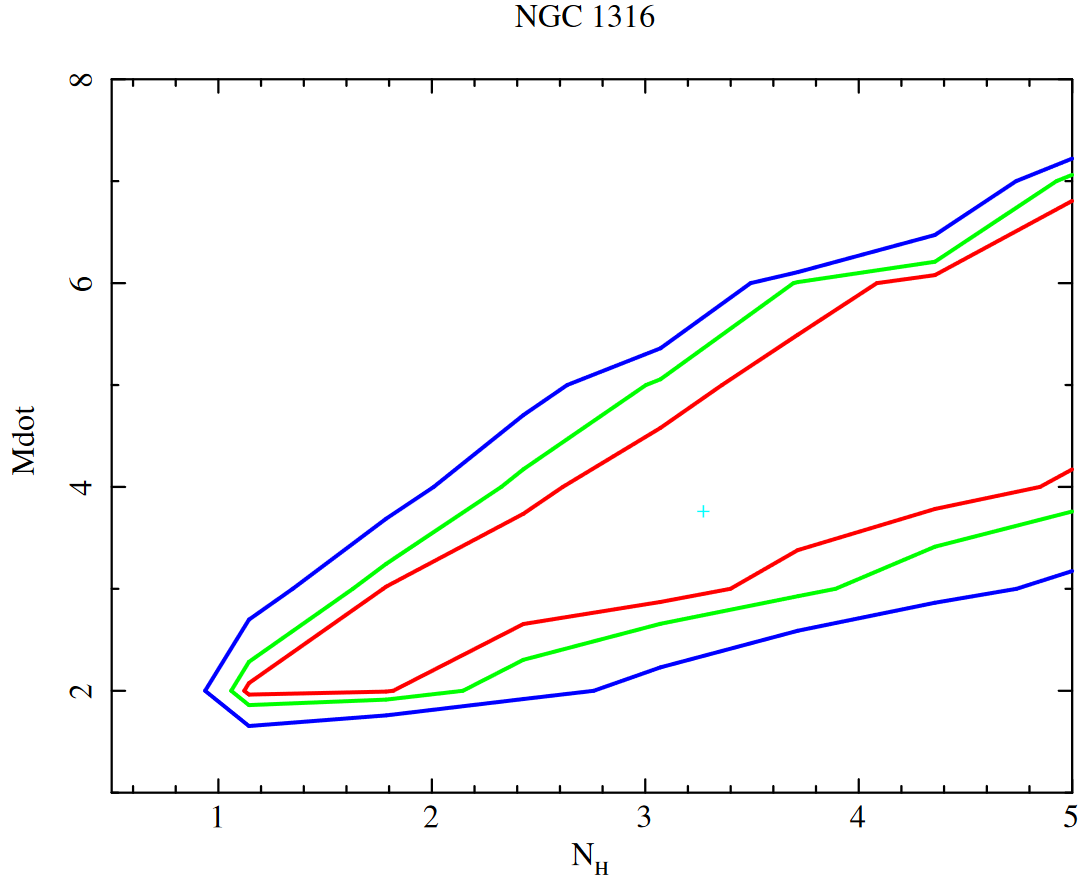}
\caption{$N_\text{H}$ vs $\dot{M}$}
\end{subfigure}
\begin{subfigure}{.47\textwidth}
\includegraphics[width=1\textwidth]{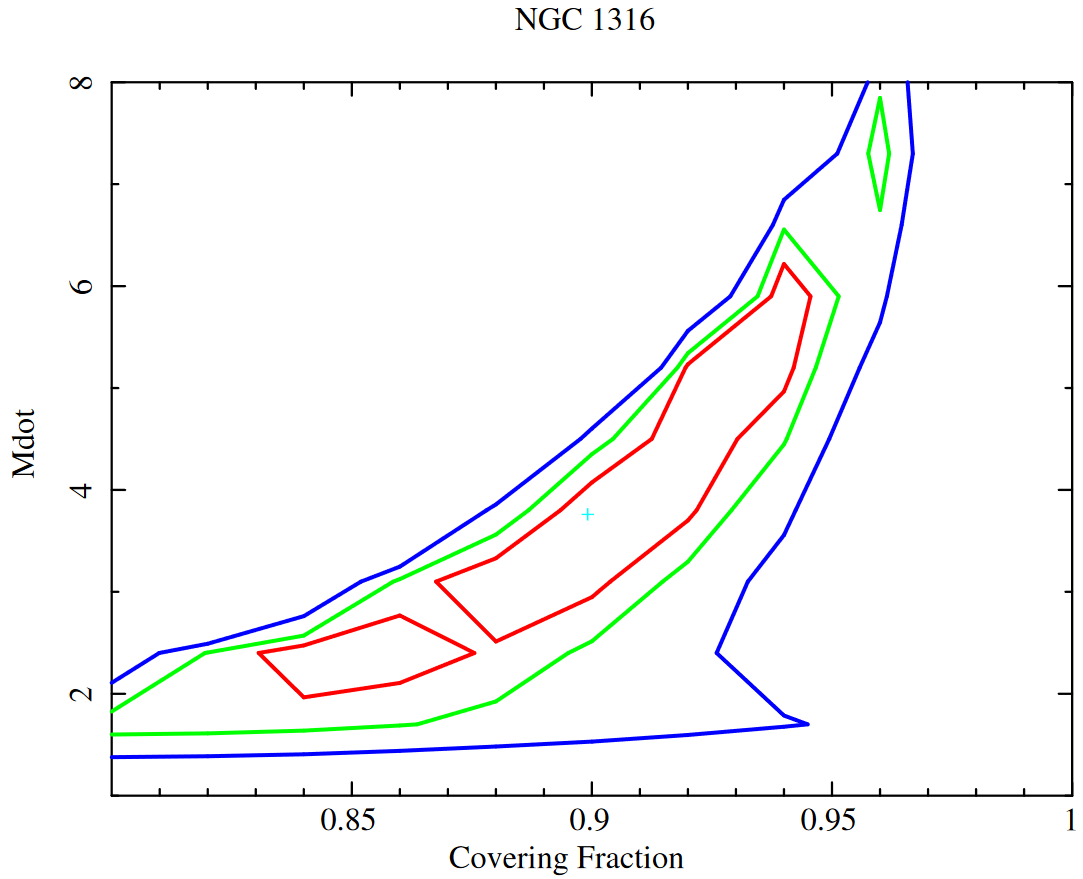}
\caption{$CFrac$ vs $\dot{M}$}
\end{subfigure}
\caption{\textbf{(a)} RGS spectrum of NGC 1316 with HCF component shown in blue and total model shown in red; \textbf{(b)} mass cooling rate in $\textup{M}_{\odot}$ yr\textsuperscript{-1} vs total column density $N_\text{H}$ in units of $10^{22}$ cm\textsuperscript{-2}; \textbf{(c)} mass cooling rate in $\textup{M}_{\odot}$ yr\textsuperscript{-1} (`\textit{Mdot}') vs \textit{Cfrac} of the HCF component. In \textbf{(b) \& (c)}, contours are at the 68\% (red), 90\% (green) and 99\% (blue) confidence intervals. The small light blue cross on each contour plot corresponds to the best-fit pair of parameters in that parameter space.}
\label{Figure4}
\end{figure}

The RGS spectra require an HCF of about 4.2 $\textup{M}_{\odot}$ yr\textsuperscript{-1}. Out of the all the galaxies in the sample, the spectrum of NGC 1316 shows the most prominent O VII lines (the peaks at $\sim$22 \AA; see \hyperref[4a]{Figure 4(a)} and \citealt{PintoEtAl2016}). This indicates that at least some even cooler  gas in the galaxy is being observed uncovered. The best-fit $CFrac$ being less than 95\% (again, somewhat unusual within this sample) supports this idea. Interestingly, the overall spectral model is completely dominated by the cooling flow component, with very little contribution from the \textsc{apec} component, indicating most of the emission from this galaxy within the RGS aperture arises from the cooling flow.

\subsection{NGC 1332}

NGC 1332 is an early-type elliptical galaxy located at $z = 0.0054$ in the Eridanus cluster. It contains an SMBH of mass $6.6 \times 10^8$ \(\textup{M}_\odot\) \citep{Barth2016}. Although there are 73 X-ray bright point sources within the galaxy, there are no detected central bright sources \citep{Humphrey1, Humphrey2}, indicating there is no unobscured AGN. 

The RGS spectra reveal an HCF of about 5.8 $\textup{M}_{\odot}$ yr\textsuperscript{-1}. A $CFrac$ of 1 suggests that the galaxy is completely covered by absorption; we constrain the unabsorbed flow to about 0.23$\ \textup{M}_{\odot}$ yr\textsuperscript{-1}. 

\begin{figure}
\centering

\begin{subfigure}{0.48\textwidth}
\includegraphics[width=\textwidth]{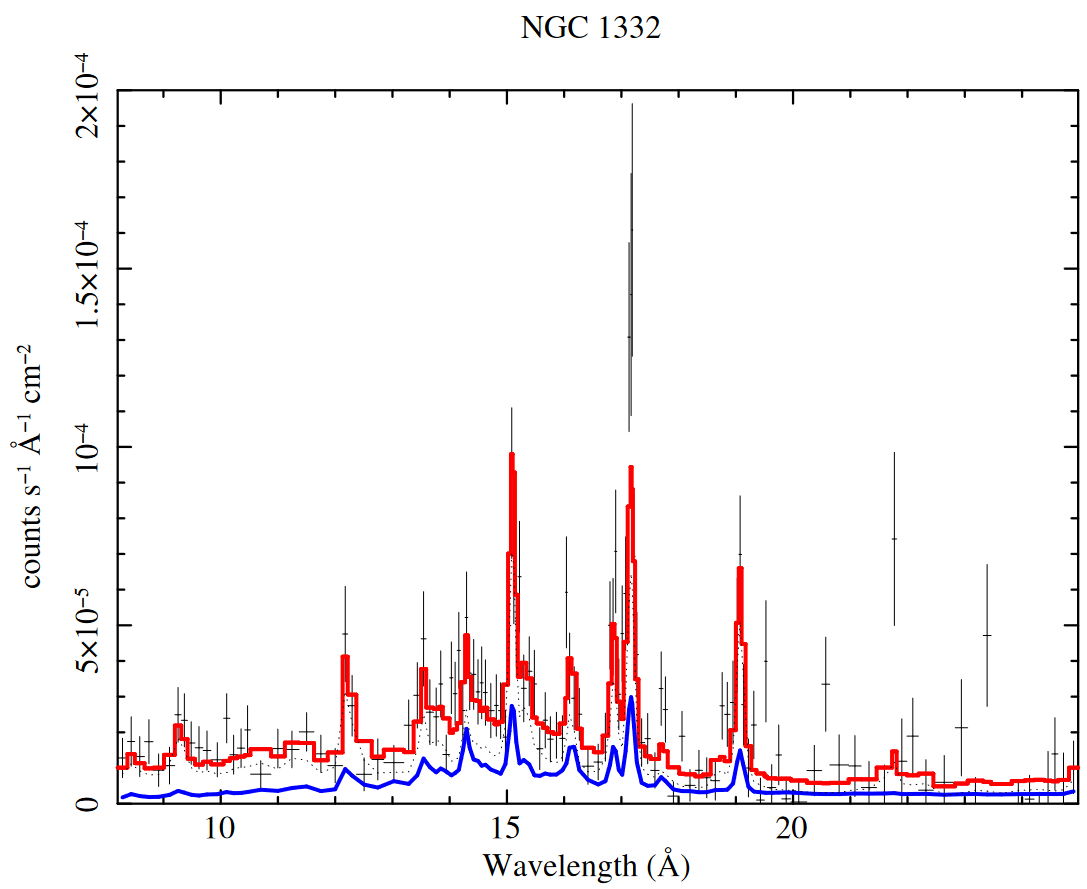}%
\caption{Spectrum}
\end{subfigure}
\begin{subfigure}{.48\textwidth}
\includegraphics[width=\textwidth]{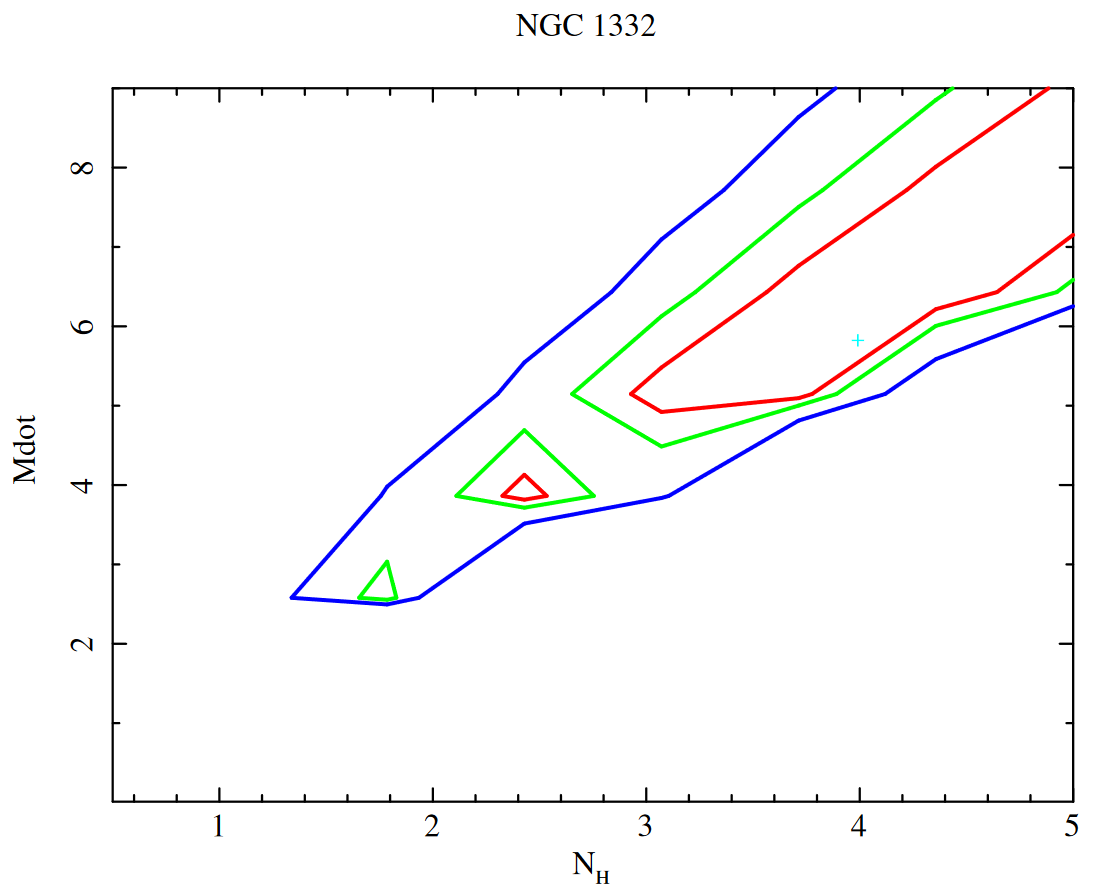}
\caption{$N_\text{H}$ vs $\dot{M}$}
\end{subfigure}
\begin{subfigure}{.48\textwidth}
\includegraphics[width=\textwidth]{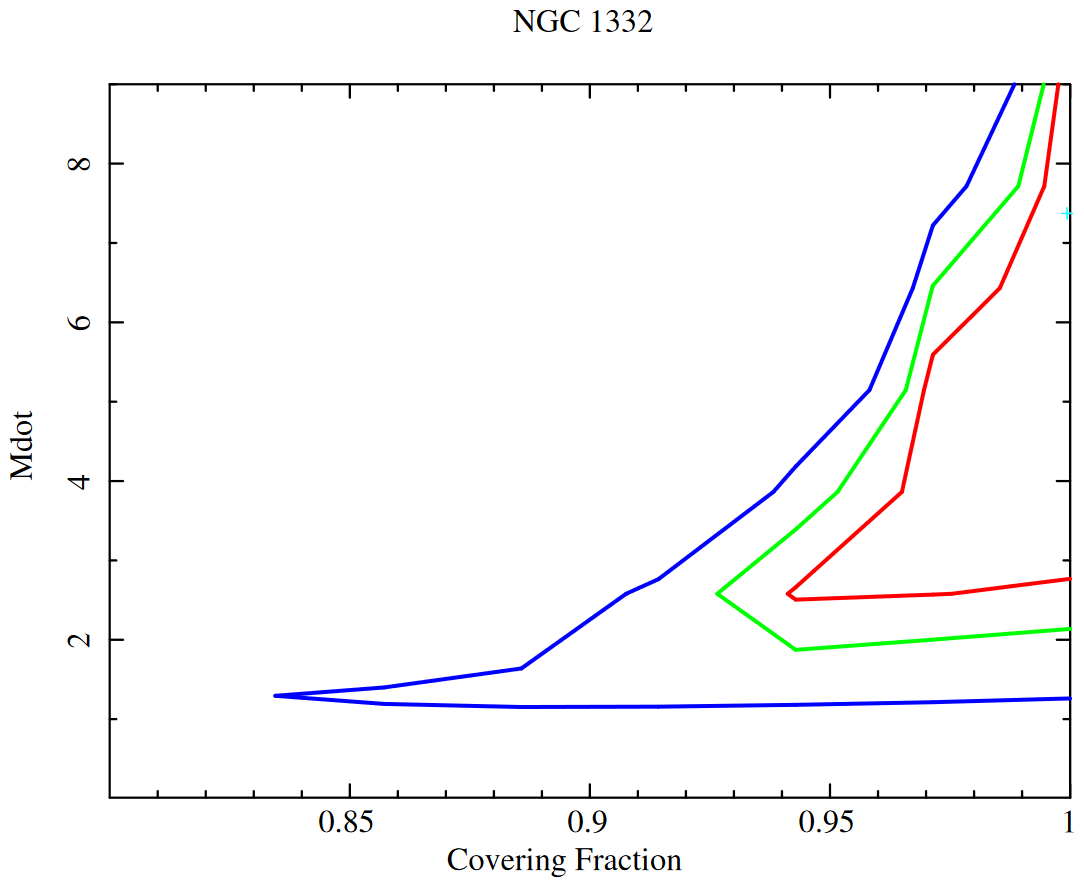}
\caption{$CFrac$ vs $\dot{M}$}
\end{subfigure}
\caption{Spectrum and contour plots of NGC 1332; other details as specified in the caption of \hyperref[Figure4]{Figure 4}.}
\label{Figure5}
\end{figure}
\begin{figure}
\centering
  \begin{subfigure}{0.48\textwidth}
  \includegraphics[width=\textwidth]{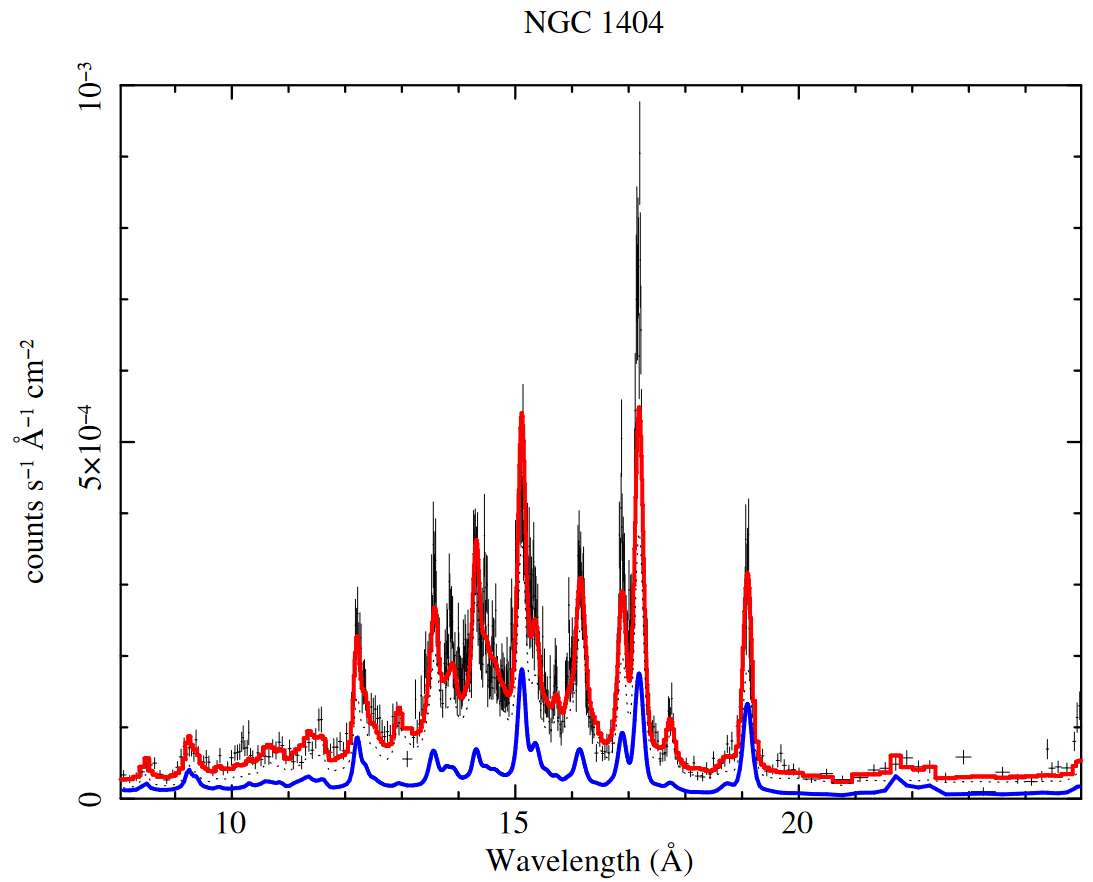}%
\caption{Spectrum}
\end{subfigure}
\begin{subfigure}{.48\textwidth}
\includegraphics[width=\textwidth]{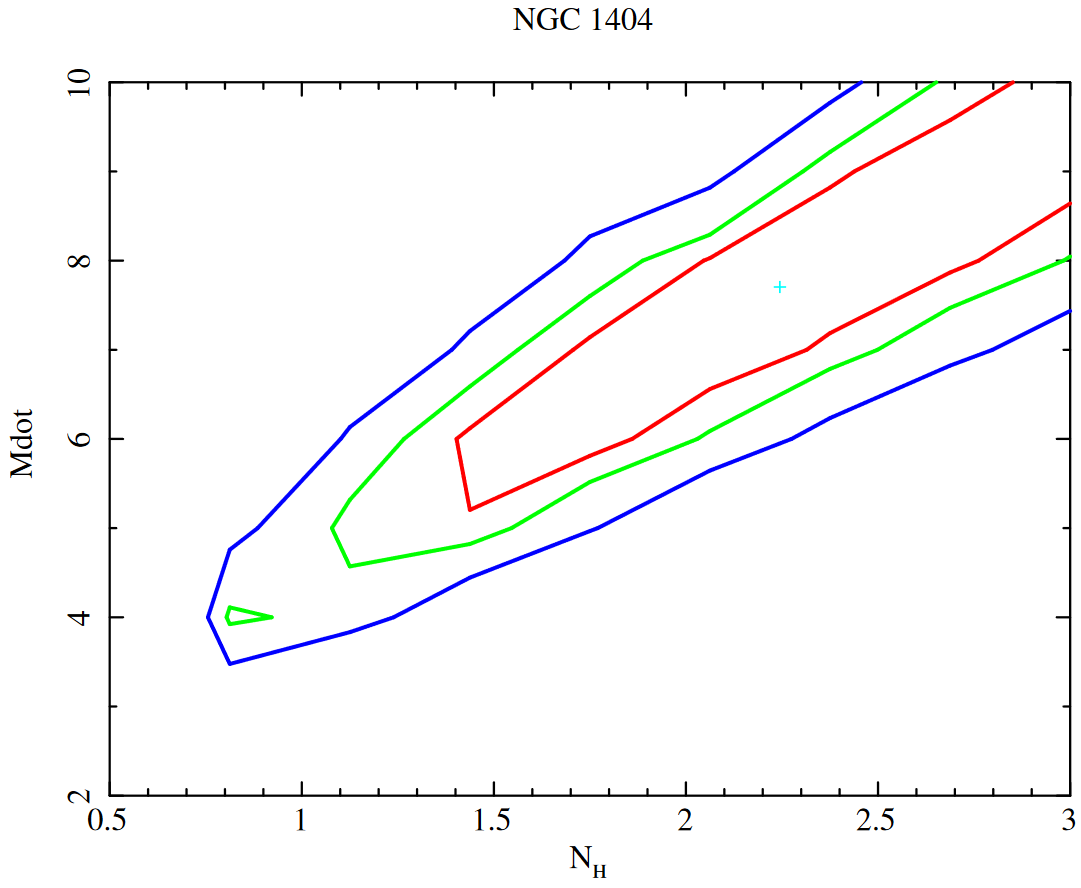}
\caption{$N_\text{H}$ vs $\dot{M}$}
\end{subfigure}
\begin{subfigure}{.48\textwidth}
\includegraphics[width=\textwidth]{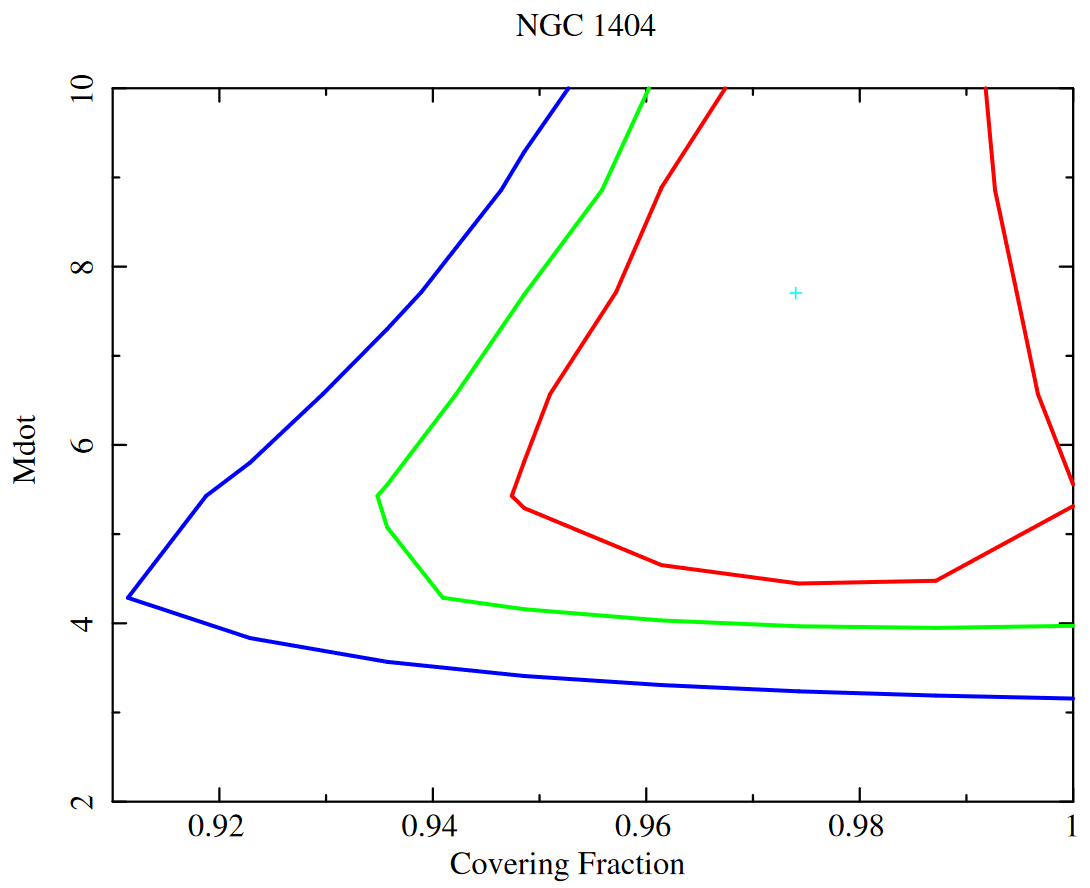}
\caption{$CFrac$ vs $\dot{M}$}
\end{subfigure}
\caption{Spectrum and contour plots of NGC 1404; other details as specified in the caption of \hyperref[Figure4]{Figure 4}.}
\label{Figure6}
\end{figure}
\begin{figure}
\centering
  \begin{subfigure}{0.48\textwidth}
  \includegraphics[width=\textwidth]{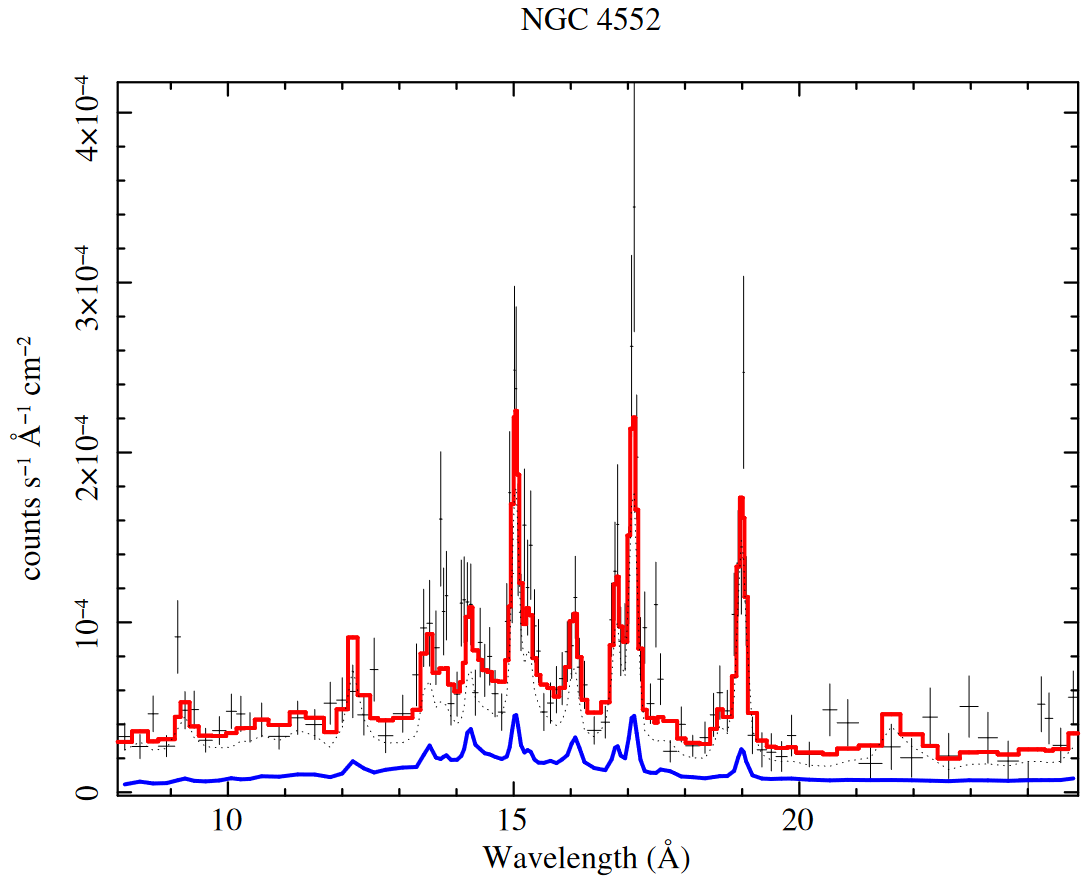}%
\caption{Spectrum}
\end{subfigure}
\begin{subfigure}{.48\textwidth}
\includegraphics[width=\textwidth]{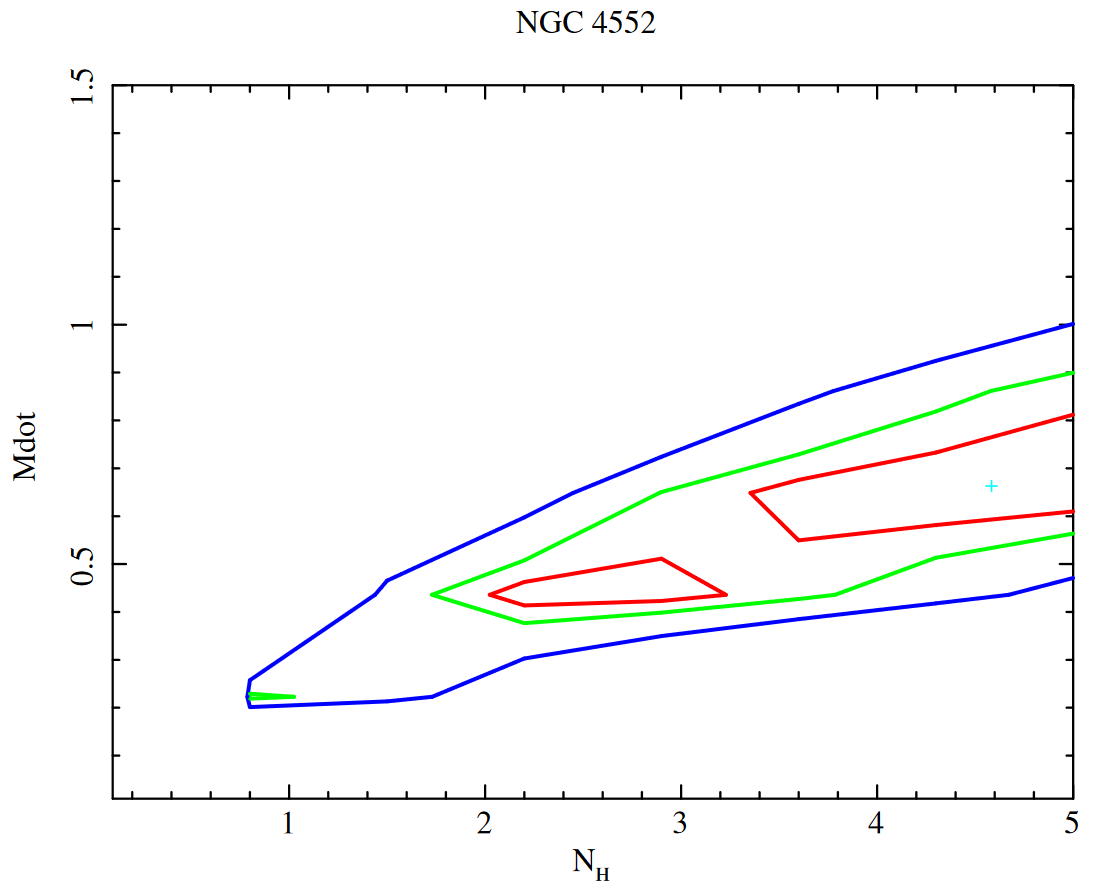}
\caption{$N_\text{H}$ vs $\dot{M}$}
\end{subfigure}
\begin{subfigure}{.48\textwidth}
\includegraphics[width=\textwidth]{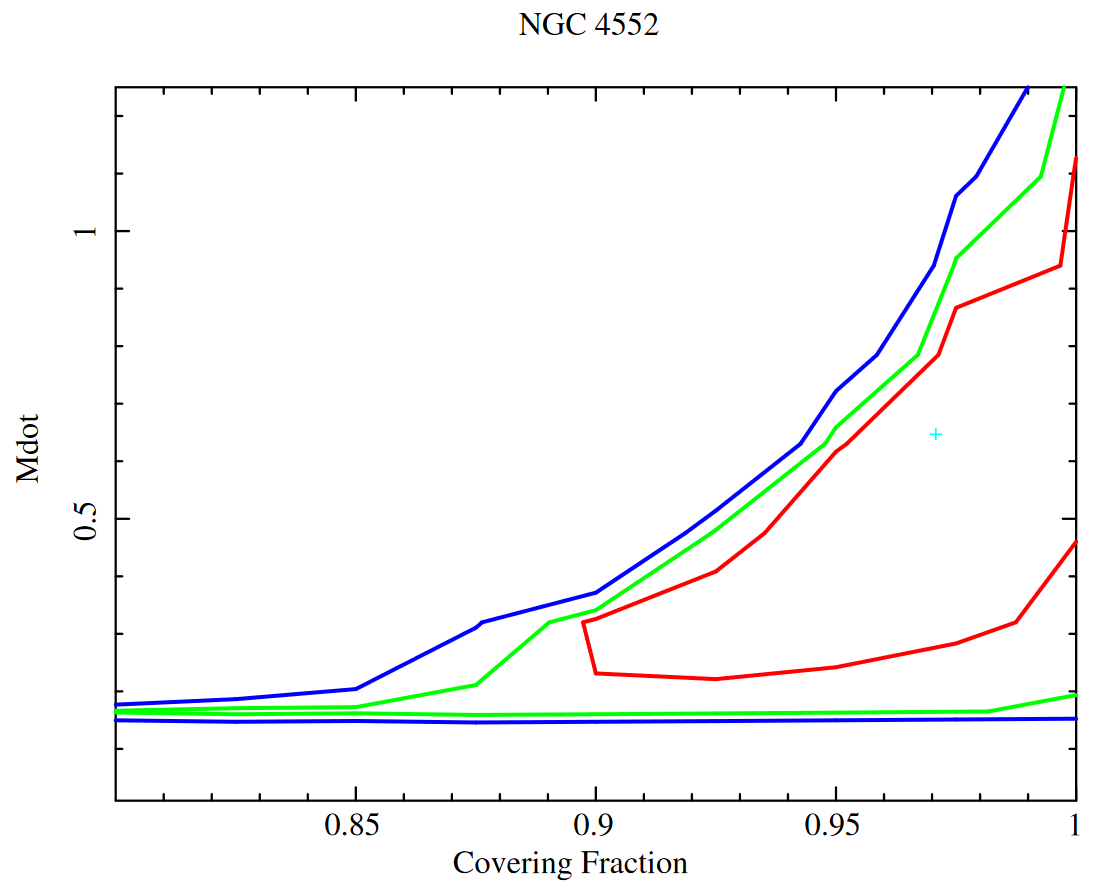}
\caption{$CFrac$ vs $\dot{M}$}
\end{subfigure}
\caption{Spectrum and contour plots of NGC 4552; other details as specified in the caption of \hyperref[Figure4]{Figure 4}.}
\label{Figure7}
\end{figure}
\begin{figure}
\centering
  \begin{subfigure}{0.48\textwidth}
  \includegraphics[width=\textwidth]{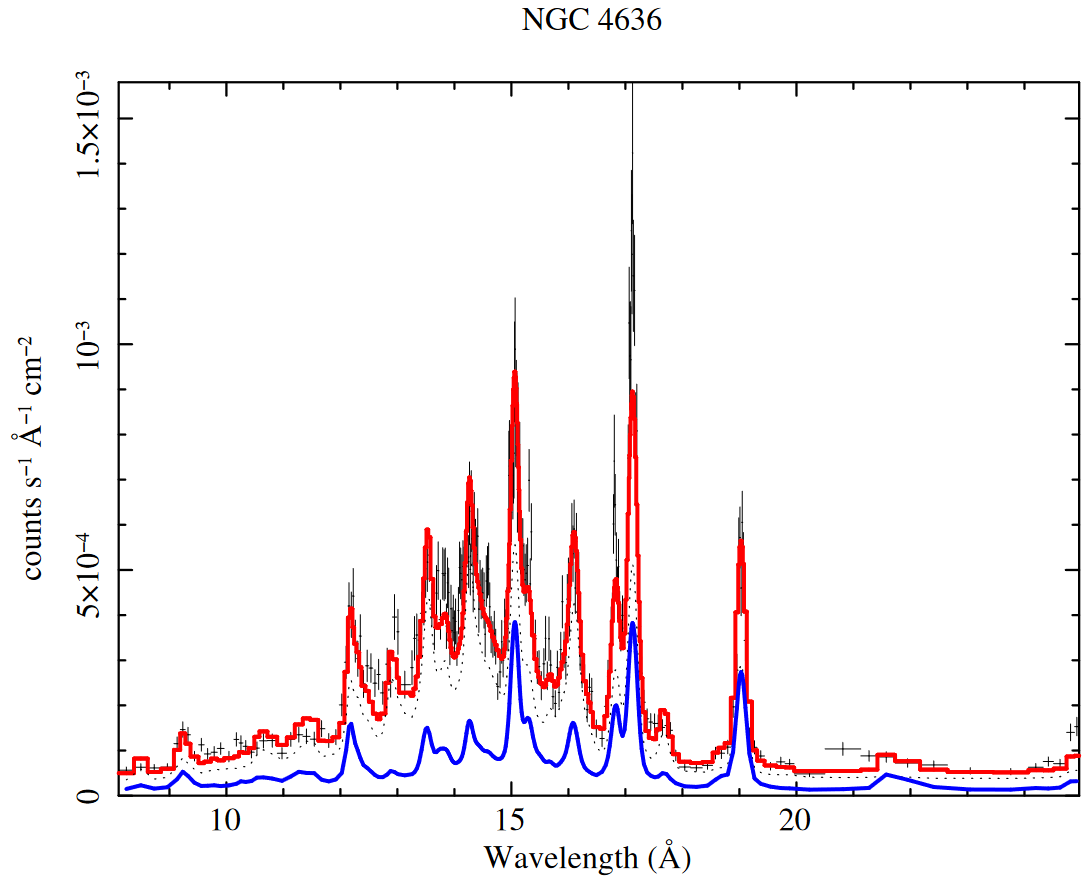}%
\caption{Spectrum}
\end{subfigure}

\begin{subfigure}{.48\textwidth}
\includegraphics[width=\textwidth]{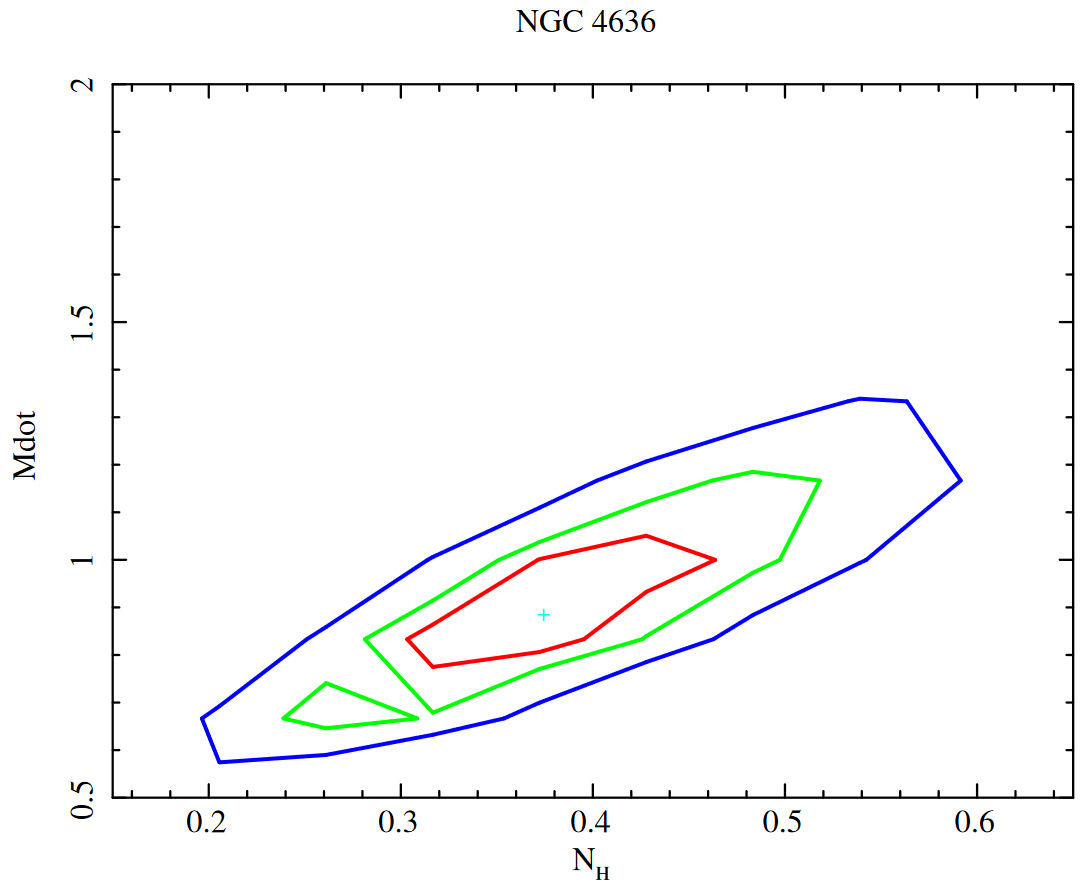}
\caption{$N_\text{H}$ vs $\dot{M}$}
\end{subfigure}
\begin{subfigure}{.48\textwidth}
\includegraphics[width=\textwidth]{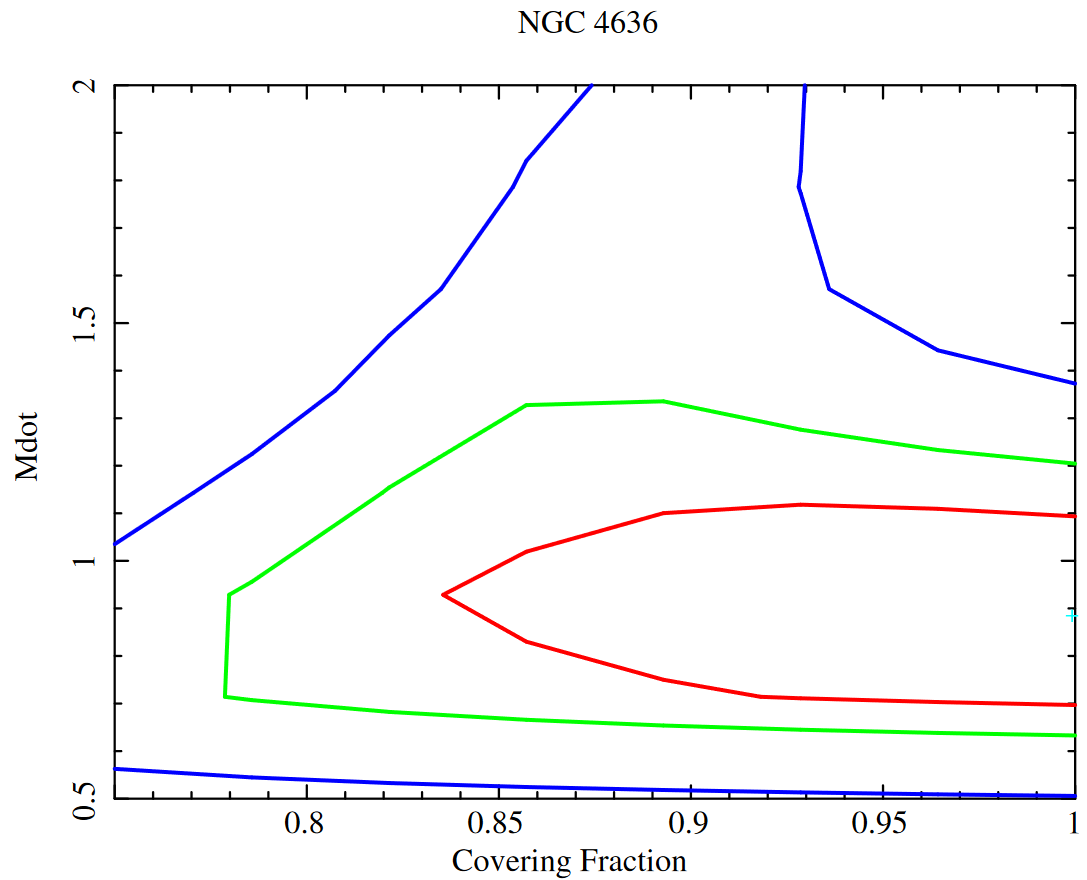}
\caption{$CFrac$ vs $\dot{M}$}
\end{subfigure}
\caption{Spectrum and contour plots of NGC 4636; other details as specified in the caption of \hyperref[Figure4]{Figure 4}.}
\label{Figure8}
\end{figure}
\begin{figure}
\centering

\begin{subfigure}{0.48\textwidth}
\includegraphics[width=\textwidth]{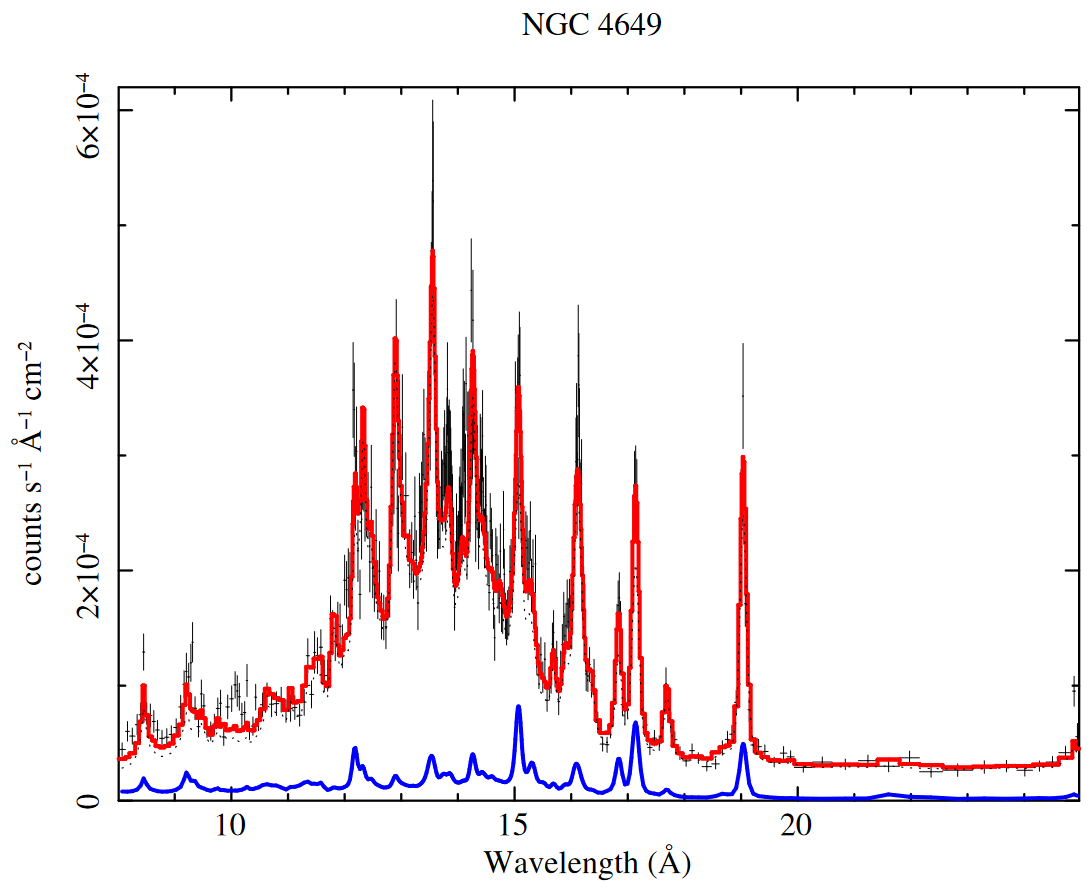}%
\caption{Spectrum}
\end{subfigure}
\begin{subfigure}{.48\textwidth}
\includegraphics[width=\textwidth]{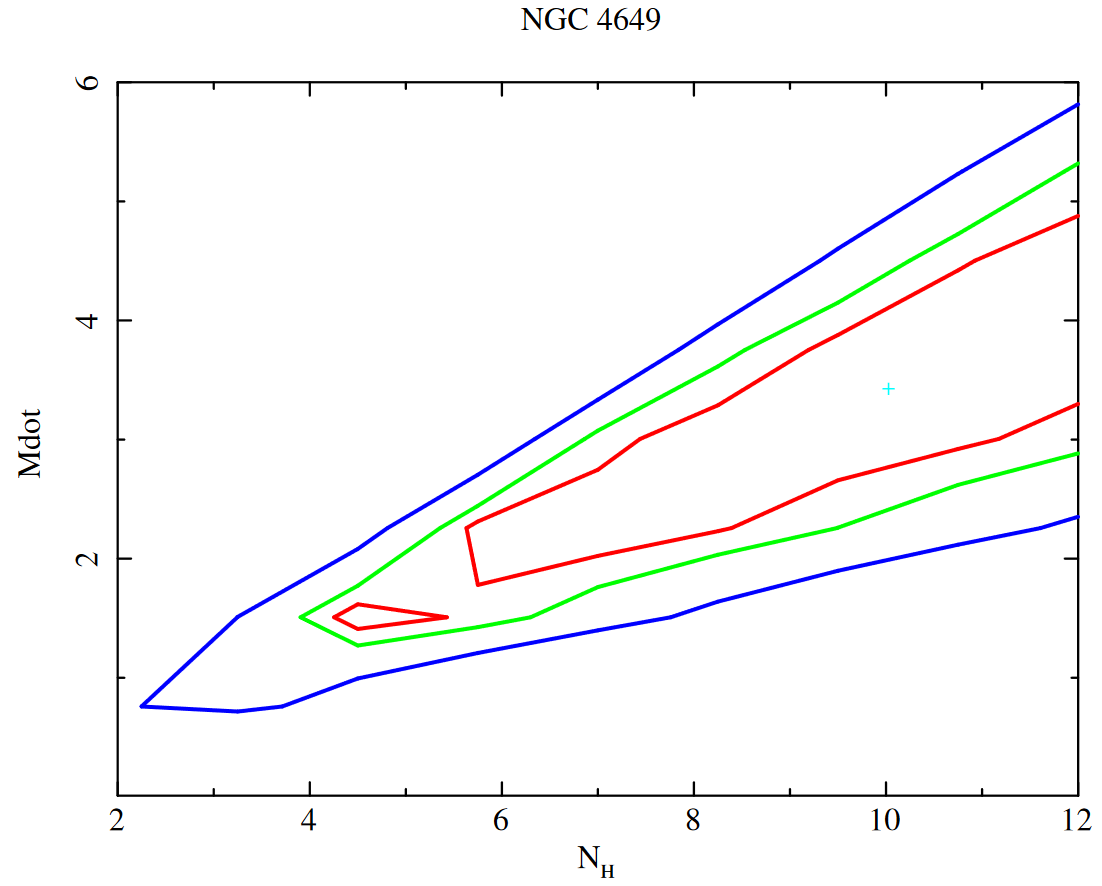}
\caption{$N_\text{H}$ vs $\dot{M}$}
\end{subfigure}
\begin{subfigure}{.48\textwidth}
\includegraphics[width=\textwidth]{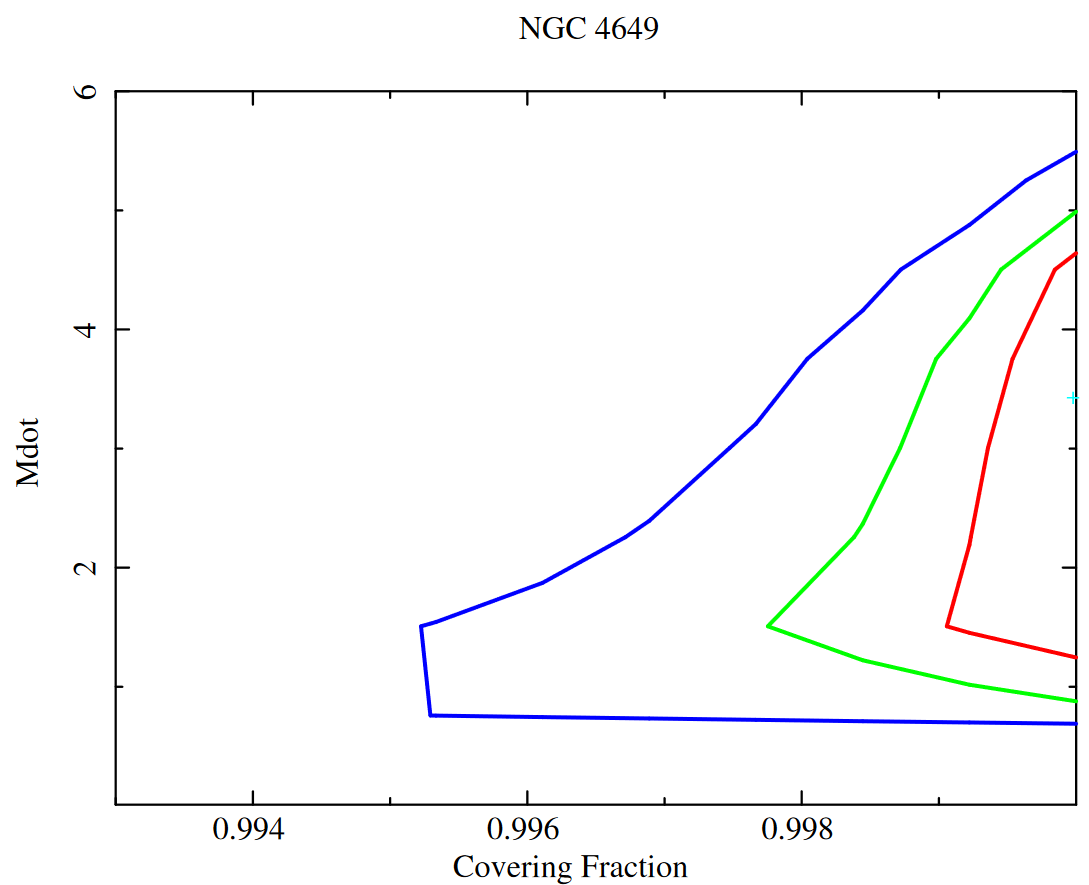}
\caption{$CFrac$ vs $\dot{M}$}
\end{subfigure}
\caption{Spectrum and contour plots of NGC 4649; other details as specified in the caption of \hyperref[Figure4]{Figure 4}.}
\label{Figure9}
\end{figure}
\begin{figure}
\centering
\begin{subfigure}[b]{0.48\textwidth}
\includegraphics[width=\textwidth]{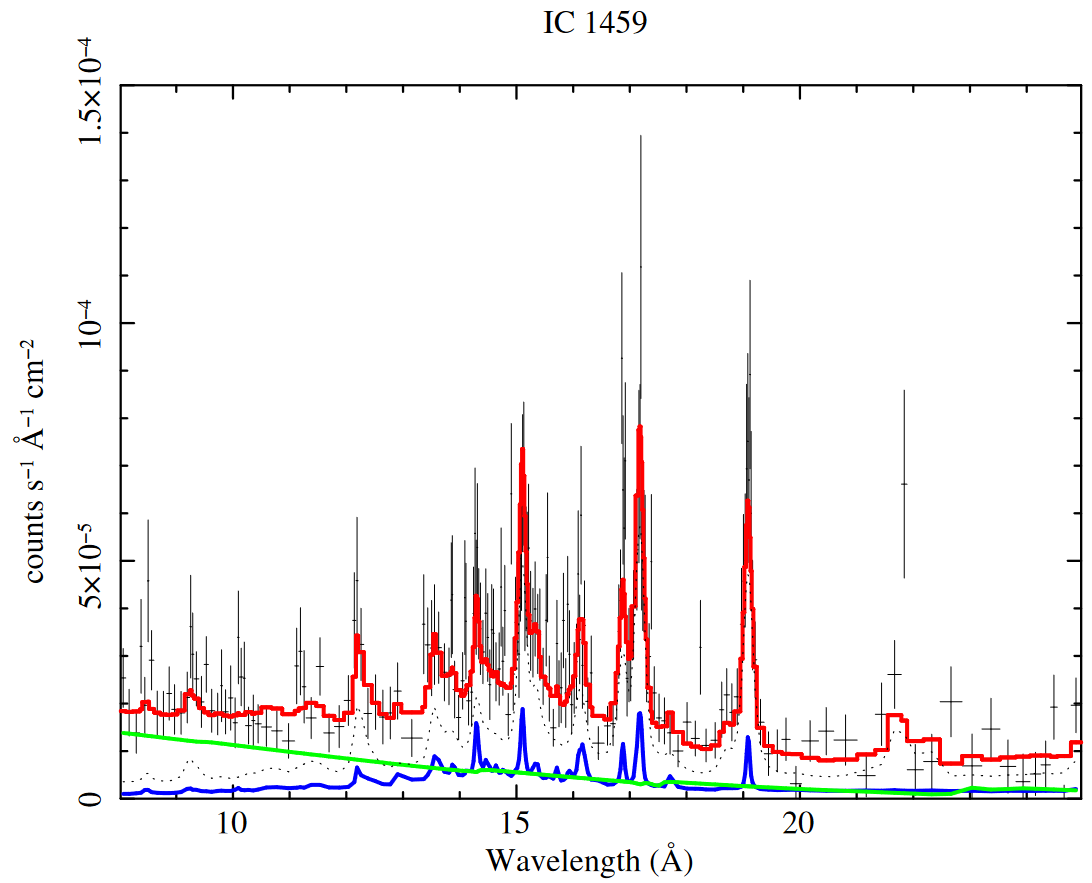}%
\caption{Spectrum}
\label{10a}
\end{subfigure}
\begin{subfigure}{.48\textwidth}
\includegraphics[width=\textwidth]{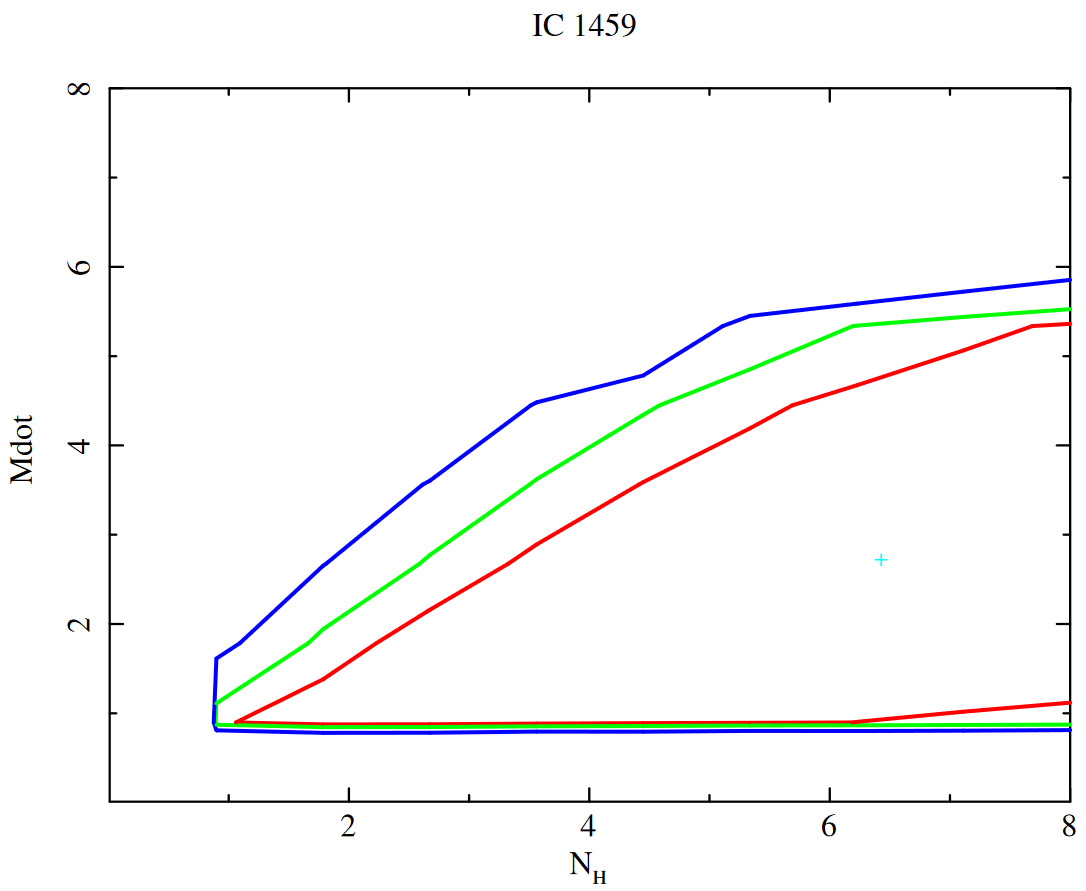}
\caption{$N_\text{H}$ vs $\dot{M}$}
\end{subfigure}
\begin{subfigure}{.48\textwidth}
\includegraphics[width=\textwidth]{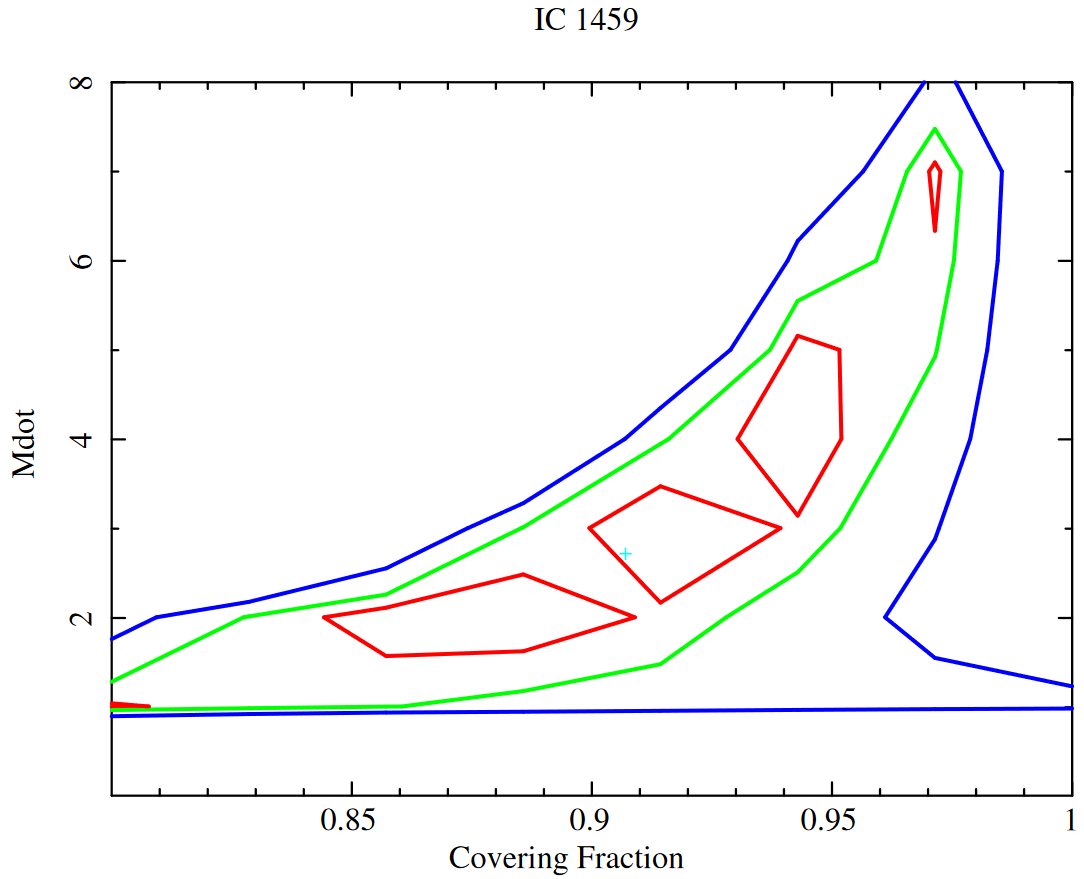}
\caption{$CFrac$ vs $\dot{M}$}
\end{subfigure}
\caption{Spectrum and contour plots of IC 1459, with other details as specified in the caption of \hyperref[Figure4]{Figure 4}. However, in \textbf{(a)}, the power-law component of the model is additionally plotted as the green line.}
\label{Figure10}
\end{figure}

\subsection{NGC 1404}

NGC 1404, at $z = 0.00649$, is also located in the Fornax cluster. Like other early-type elliptical galaxies, it contains a large population of globular clusters \citep{Forbes1998}. However, it is thought that it may have lost many of its globular clusters due to gravitational interactions with the nearby NGC 1399, which is the brightest galaxy in the Fornax cluster \citep{Bekki2003}. Additionally, ram pressure stripping caused by NGC 1404's motion through the intracluster medium is removing some of the galaxy's hot gas and leaving a large tail \citep{Machacek2005}. While X-ray observables have been used to probe the ICM around this galaxy, there is no bright nuclear source \citep{Su2017}.

\cite{Bregman2001} found that O VI lines were not detected in the spectrum of this galaxy, which constrained the estimated cooling flow rate to 0.3 $\textup{M}_{\odot}$ yr\textsuperscript{-1}. This value represents a lower bound on the mass cooling flow rate, as they did not account for absorption in their calculation. The RGS spectra allow an HCF of 7.7 $\textup{M}_{\odot}$ yr\textsuperscript{-1}. The flow is almost completely covered by absorption, allowing a maximum unabsorbed cooling rate of 0.55 $\textup{M}_{\odot}$ yr\textsuperscript{-1}. 

\subsection{NGC 4552}

NGC 4552 (also known as \textit{Messier 89}) is a member of the Virgo cluster, whose centre is about 16.5 Mpc away \citep{Mei_2007}; this means the measured redshift of  $z = 0.00113$ is unreasonably small. This is a consequence of its motion within the cluster, as NGC 4552 in particular is moving towards the observer \citep{Machacek2006strip}. When calculating FIR luminosity in XSPEC (see \hyperref[Further]{Section 6}), we adjusted to $H_0 = 20$ km s\textsuperscript{-1} Mpc\textsuperscript{-1} to ensure the distance to NGC 4552 was corrected to match redshift-independent measurements (see \hyperref[Table2]{Table 2}). This is the only galaxy within the sample for which such a correction was necessary.

NGC 4552 has large surrounding structures of gas and dust, and is likely to have experienced multiple mergers \citep{Janowiecki_2010}. It also contains a central supermassive black hole of $4.8 \times 10^8$ \(\textup{M}_\odot\) \citep{Graham_2008}. Studies of structures of hot gas in the galaxy indicate that there was an outburst in the galactic nucleus a few million years ago \citep{Machacek2006}, which suggests NGC 4552 may have once been an active quasar or radio galaxy.  Similar to NGC 1404, NGC 4552 is also experiencing ram-pressure stripping as it moves through the intracluster medium \citep{Machacek2006strip}.

An HCF of 0.65 $\textup{M}_{\odot}$ yr\textsuperscript{-1} is allowed by the spectra; the corresponding unabsorbed flow is 0.034 $\textup{M}_{\odot}$ yr\textsuperscript{-1}. It has the lowest HCF rate found in this sample of galaxies.

\subsection{NGC 4636}

NGC 4636 is an elliptical galaxy at $z = 0.00313$, which is a member of the NGC 4753 Group of galaxies, one of the Virgo II groups strung out from the edge of the Virgo Supercluster.

This galaxy has an AGN, which is classified as a low-ionisation nuclear emission-line region or LINER \citep{Ho1997}. However, the SMBH powering this activity is actually lower in mass than those of other galaxies in this sample, being only $7.9\times 10^7 \ \textup{M}_{\odot}$ \citep{Merritt2001}. The bright, dense core of this galaxy contains a number of X-ray point sources, one of which is coincident with the radio nucleus, and could correspond to X-ray emission from the SMBH \citep{Baldi2009}. However, the low luminosity of this source suggests it could be due to a low-mass X-ray binary (LMXB) instead.

NGC 4636 is one of the most luminous nearby elliptical galaxies when observed in X-rays \citep{Forman1985}. There are several bubble-like features in the halo, likely the result of shocks generated by AGN jets \citep{Baldi2009}. The bubbles may be of different ages, generated by different AGN outbursts; evidence for this lies in radio-emitting plasma being present in some cavities but not others \citep{Giacintucci2011}. The core of NGC 4636 is X-ray bright, with a cavity that is likely generated by a radio jet at 1.4 GHz \citep{Allen2006}. This points to a scenario in which some gas may be currently outflowing in the central kpc of the galaxy \citep{Temi2018}.

Fitting the RGS spectra allows for an HCF of $\sim$1 $\textup{M}_{\odot}$ yr\textsuperscript{-1}; the flow is likely completely covered by absorption. \cite{Bregman2001} identified O VI lines in the spectrum of this galaxy, associated with gas at temperatures $\sim$10\textsuperscript{5} K and hence with effective gas cooling. They determined a cooling flow rate of $0.43 \pm 0.06 \ \text{M}_{\odot}$ yr\textsuperscript{-1}, but did not account for absorption. Including absorption in their analysis could bring their mass flow rate estimate in agreement with the value estimated from our HCF model.
 
 Out of all of the galaxies in the sample, NGC 4636 has the second-lowest calculated HCF, which may be because of AGN-driven gas outflows mentioned above. Interestingly, the spectral lines of NGC 4636 are significantly stronger than in any of the other galaxies in this sample. This correlates with the galaxy having an X-ray bright core, and with NGC 4636 having the lowest value of $N_\text{H}$ in the sample. 

\subsection{NGC 4649}

NGC 4649 (also known as \textit{Messier 60}) at $z = 0.0037$ is also a member of the Virgo cluster. It forms a pair with NGC 4647, which it may just be beginning to tidally interact with \citep{deGrijs2006}.

It contains a supermassive black hole of mass 4.5 $\times \ 10^9$ \(\textup{M}_\odot\) \citep{Shen_2010}; while it is currently inactive, inspection of the galaxy's structure using X-rays shows structure caused by jets produced by the SMBH's past active periods \citep{Shurkin2008}. There is an X-ray point source at the centre of the galaxy which is consistent with its optical nucleus, with luminosity $L_\text{X}\sim 10^{38}$ erg s\textsuperscript{-1} \citep{Baldi2009}. This could be from either an AGN or an LMXB, as is the case for NGC 4636. \cite{2022Temi} found that this galaxy is apparently devoid of dust, which explains the lack of obvious absorption regions in HST images of the galaxy (see \hyperref[Figure11]{Figure 11(f)}).

The spectra allow an HCF of up to $3.4 \ \textup{M}_{\odot}$ yr\textsuperscript{-1}; to do so, a total intrinsic absorption column density of $N_{\text{H}} = 10.06 \times 10^{22}$ cm\textsuperscript{-2} was required. This is a significantly greater $N_{\text{H}}$ than was necessary for any of the other galaxies studied here, which is unusual. The unabsorbed mass cooling flow allowed in this galaxy is negligible.

\subsection{IC 1459}\label{Section3.7}
IC 1459 is a giant elliptical galaxy at $z = 0.00601$. Its formation history likely included the merger of several smaller galaxies with the elliptical galaxy \citep{Tal2009}. It has an active nucleus and, like NGC 4636, has been classified as a LINER \citep{Phillips1986}. The mass of the central black hole has been calculated using stellar kinematics to be  2.6 $\times \ 10^9$ \(\textup{M}_\odot\) \citep{Cappellari2002}, although other mass measurement methods have yielded slightly different values (see \citealt{Verdoes2000}, for example). The circumnuclear gas is found to be kinetically disturbed, which may be indicative of an outflow \citep{Ricci2015}.

The galaxy is radio-loud, with a GHz peaked spectrum \citep{Tingay2015}; it also has X-ray emission. Chandra observations of the supermassive black hole in the nucleus of IC 1459 show a weak unabsorbed nuclear X-ray source. \cite{Fabbiano2003} found that this source could be described by a power-law, with normalisation $2.1 \times 10^{-4}$ and index $1.88$. This unabsorbed X-ray radiation contributed significantly to the observed X-ray continuum, reducing the line abundance. To account for this source, a power-law component was added to the model, so that the overall model in XSPEC became \textsc{tbabs(gsmooth*apec+tbabs*powerlaw+gsmooth(partcov* mlayerz)mkcflow)}. The power-law has an additional absorption of $0.29 \times 10^{22}$ cm\textsuperscript{-2} applied to it. All parameters associated with the power-law and its abundance were frozen at the values specified by \cite{Fabbiano2003} prior to fitting.

The spectra allow an HCF of up to 2.7 $\textup{M}_{\odot}$ yr\textsuperscript{-1}, and an unabsorbed mass flow of up to 0.3 $\textup{M}_{\odot}$ yr\textsuperscript{-1}. The covering fraction of 0.91 is also consistent with part of the AGN activity being being observed uncovered, and the small O VII lines at $\sim$22 \AA \ (see \hyperref[10a]{Figure 10(a)}) indicate that at least some low-temperature gas is being observed directly.

\section{Hubble Space Telescope Images}
\begin{figure*}
% first row: 3 subfigures
\begin{subfigure}{0.3\textwidth}
   \includegraphics[width=\linewidth]{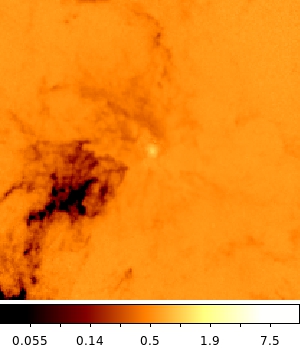}
   \caption{\textbf{NGC 1316}; $1.29 \times 1.29$ kpc} \label{11a}
\end{subfigure}
\hspace*{\fill}
\begin{subfigure}{0.3\textwidth}
   \includegraphics[width=\linewidth]{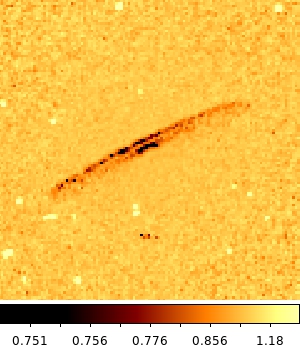}
   \caption{\textbf{NGC 1332}; $0.69\times 0.69$ kpc} \label{11b}
\end{subfigure}
\hspace*{\fill}
\begin{subfigure}{0.3\textwidth}
   \includegraphics[width=\linewidth]{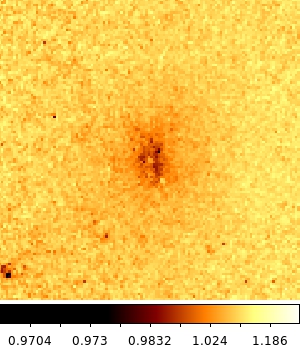}
   \caption{\textbf{NGC 1404}; $0.83 \times 0.83$ kpc} \label{11c}
\end{subfigure}

% 2nd row: 3 more subfigures
\bigskip
\begin{subfigure}{0.3\textwidth}
   \includegraphics[width=\linewidth]{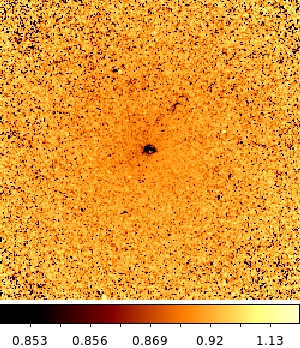}
   \caption{\textbf{NGC 4552}; $0.34 \times 0.34$ kpc} \label{11d}
\end{subfigure}
\hspace*{\fill}
\begin{subfigure}{0.3\textwidth}
   \includegraphics[width=\linewidth]{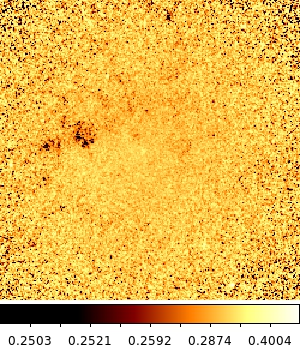}
   \caption{\textbf{NGC 4636}; $0.94 \times 0.94$ kpc} \label{11e}
\end{subfigure}
\hspace*{\fill}
\begin{subfigure}{0.3\textwidth}
   \includegraphics[width=\linewidth]{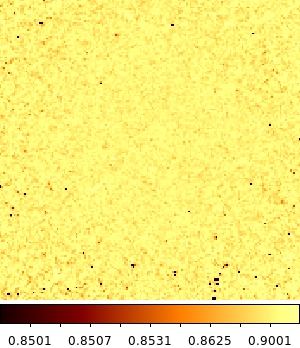}
   \caption{\textbf{NGC 4649}; $0.72 \times 0.72$ kpc} \label{11f}
\end{subfigure}

\bigskip
\hspace*{\fill}
\begin{subfigure}{0.3\textwidth}
   \includegraphics[width=\linewidth]{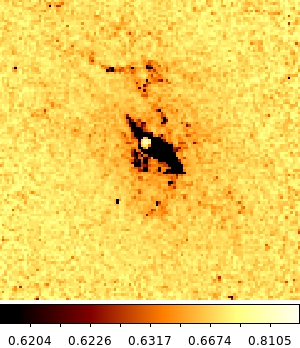}
   \caption{\textbf{IC 1459}; $0.90 \times 0.90$ kpc} \label{11g}
\end{subfigure}%
\hspace*{\fill}

\caption{HST images of the centres of the galaxies studied in this paper, prepared by combining a red and blue image of each galaxy. The images were scaled to have the same maximum brightness prior to combination. The centre of each image corresponds to the galactic centre \& dark regions indicate absorption. The dimensions of each image in kpc, specified in the relevant sub-caption, were calculated using the redshift-independent distance measurements listed in \hyperref[Table1]{Table 1}. The legend at the bottom of each image is specified in arbitrary flux units. All image data is sourced from the Hubble Legacy Archive.}
\label{Figure11}
\end{figure*}

Evidence for cool gas in elliptical galaxies can also be found by studying them in optical wavelengths. \hyperref[Figure11]{Figure 11} shows Hubble Space Telescope (HST) images of absorption near the centres of several elliptical galaxies, prepared by combining a red (814 nm) and blue (450 nm, 547 nm or 555 nm, depending on data availability from the Hubble Legacy Archive) image of each galaxy. Absorption by dust is stronger for the blue wavelength than the red. The red and blue images were normalised to the same central intensity, then the blue (higher-energy) image was divided by the red image. Combining the images in this way brings absorption into stark relief as the dark regions in each resulting image.

NGC 1316, NGC 1332 \& IC 1459 (see \hyperref[11a]{Figure 11(a)}, \hyperref[11b]{(b)} \& \hyperref[11g]{(g)}) in particular demonstrate large, structured absorption regions near the galactic centre. On the other hand, note that NGC 4649 (\hyperref[11f]{Figure 11(f)}) lacks any distinct regions of absorption at optical wavelengths. Since it has the highest value of $N_\text{H}$ in the sample, any absorption regions could be deeply buried. It could be that optical images alone are insufficient to search for dust; see \hyperref[Further]{Section 6} for a comparison with FIR dusty emission. NGC 1404 \& NGC 4552 (see  \hyperref[11c]{Figure 11(c)} \& \hyperref[11d]{(d)}) have smaller, less structured absorption around the galactic centres. NGC 4636 (see \hyperref[11e]{Figure 11(e)}) has a small region of absorption, but it is actually slightly above and to the left of the galactic centre in this image.

The absorption observed is caused by dust. If dust is in a hot gas environment, it is quickly broken up by sputtering \citep{Draine1979}. So the presence of the dust indicates the surrounding gas is cold. Consequently, we infer that the absorption seen in \hyperref[Figure11]{Figure 11} is evidence for cold and cooling gas near the centres of the galaxies under investigation.

\section{Correcting for Point Sources}\label{Correction}

\begin{figure}
     \centering
     \includegraphics[width=0.48\textwidth]{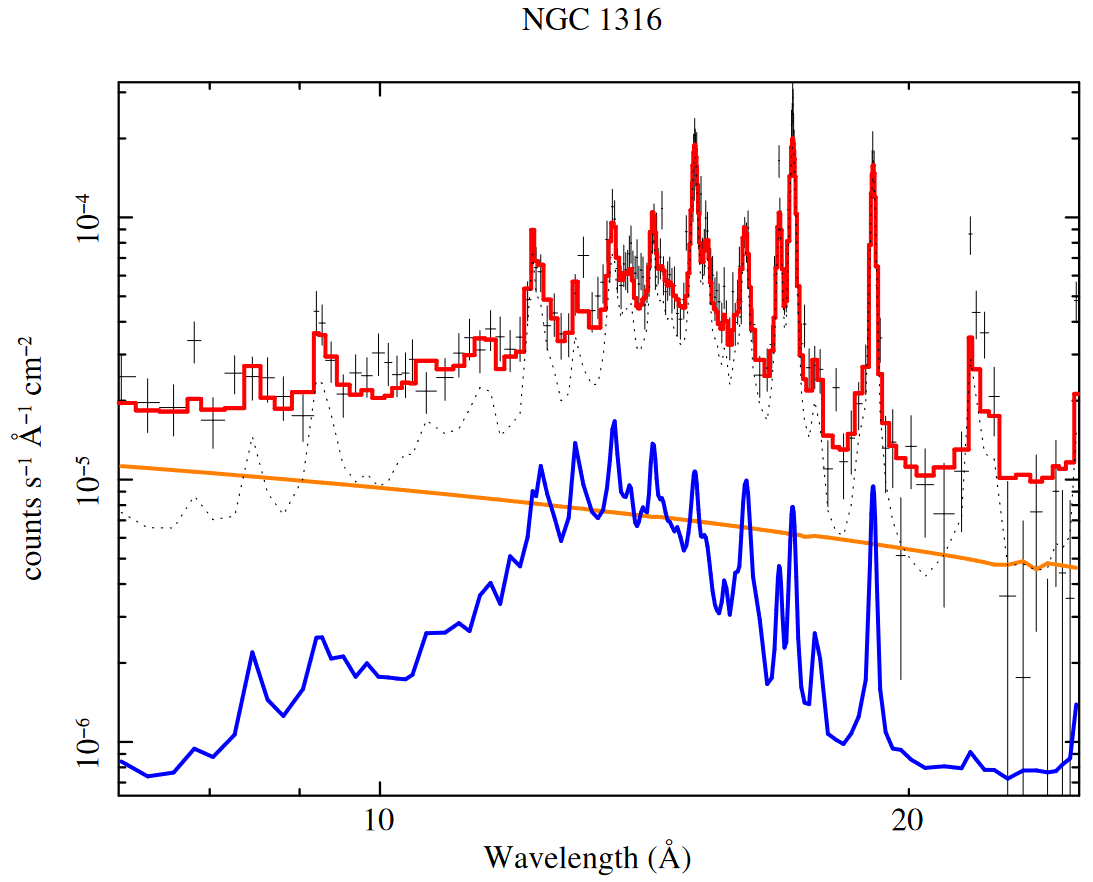}
    \label{Figure 12}
    \caption{Spectral model of NGC 1316, now including the \textsc{bremss} component of the model as the orange line. The cooling flow component and overall model are the blue and red lines respectively.}
\end{figure}

{\renewcommand{\arraystretch}{1.3}
\begin{table*}
\normalsize
\centering
 \begin{tblr}{c c c c c c c c }
 \hline
 Target & Bremss Norm & $kT$ & $Z$ & APEC Norm & $CFrac$ & $N_\text{H}$& $\dot{M}$\\ 
 \hline
  & $10^{-5}$ & keV & $Z_{\odot}$& $10^{-4}$ & &$10^{22}$ cm \textsuperscript{-2} & $M_\odot$ yr\textsuperscript{-1} \\
 \hline
 NGC 1316 & 14.1 & 0.90 & $0.53_{-0.17}^{+0.46}$ & 0.56 & 0.906 & 5.0 & $3.2_{-2.7}^{+1.0}$\\
 NGC 1332 & 9.7 & 0.61 & $0.21_{-0.08}^{+0.11}$ & 2.4 & 1 & 5.0 & $2.9_{-2.8}^{+3.8}$\\
 NGC 1404 & 9.6 & 0.69 & $0.22_{-0.06}^{+0.47}$ & 18.5 & 0.97 & 2.3 & $7.4_{-1.8}^{+0.4}$\\
 NGC 4552 & 29.3 & 0.58 & $0.13_{-0.07}^{+0.10}$ & 9.1 & 1 & 4.9 & $0.56_{-0.21}^{+0.45}$\\
 NGC 4636 & 2.6 & 0.75 & $0.28_{-0.18}^{+0.22}$ & 30.1 & 1 & 0.33 & $0.80_{-0.56}^{+0.7}$\\
 NGC 4649 & 29.7 & 0.88 & $0.63_{-0.13}^{+0.20}$ & 12.3 & 1 & 11.4 & $0.80_{-0.79}^{+0.99}$\\
 IC 1459 & 3.6 & 0.72 & $0.16_{-0.07}^{+0.14}$ & 1.1 & 0.81 & 10.0 & $1.3_{-1.2}^{+3.9}$\\
 \hline
\end{tblr}
\caption{Table of model parameters obtained by refitting the models with an additional 7.3 keV \textsc{bremss} model component, intended to account for the LMXB contribution to the spectra. Note that abundances, while still low, have at least increased slightly compared to the previous models. 90\% confidence intervals are specified only for $Z$ and $\dot{M}$.}
\label{Table5}
\end{table*}
}
Chandra X-ray Observatory studies of these objects (mentioned in \hyperref[Results]{Section 3}) have found that they each contain multiple bright point sources. \cite{Irwin2003} and others have argued that such point sources are  low mass X-ray binaries (LMXBs), which are charactertised by a featureless bremsstrahlung spectrum of temperature $7.3$ keV. Here, we make a simple correction to the model spectra presented in the previous Section, by including such a component. The \textsc{bremss} component of the model, which is also subject to Galactic absorption, is shown as the orange line in \hyperref[Figure 12]{Figure 12}. We extended the spectra down to 7 \AA \ in order to obtain a better continuum fit. The results are summarised in \hyperref[Table5]{Table 5}.

Including the continuum emission from the LXMBs has caused the abundances to increase, while the values of $\dot{M}$ generally remain generally similar to those found previously. The abundances found here are broadly compatible with the results of \cite{Lakhchaura18}, which identified abundances of one-third of the solar value for most of the galaxies in their sample.

The integrated luminosity associated with the \textsc{bremss} model component was calculated, and compared with the relevant total X-ray luminosities of point sources in 4 of our galaxies in a sample studied by \cite{KimFabbiano2004}. While we would not expect complete agreement in values due to different observation regions between our study and theirs, the luminosities are similar and indicate that the point source correction is appropriate. 

\section{Discussion}\label{Further}

\begin{table}
\footnotesize
\centering
 \begin{tabular}{c c c c} 
 Target & $\log(M_{*} / M_\odot)$ & $\dot{M_{*}}$ & $\dot{M}$\\ 
 \hline
 &  & $M_\odot \ \text{yr}^{-1}$ & $M_\odot \ \text{yr}^{-1}$\\
 \hline
 NGC 1316 & 11.84 & 2.9 & 4.2\\
 NGC 1332 & 11.27 & 0.78 & 5.8\\
 NGC 1404 & 11.10 & 0.61 & 7.7\\
 NGC 4552 & 11.42 & 1.0 & 0.65\\
 NGC 4636 & 11.84 & 2.9 & 0.89\\
 NGC 4649 & 11.64 & 1.5 & 3.4\\
 IC 1459 & 11.60 & 1.7 & 2.7\\
 \hline
 
 \end{tabular}
 \caption{Comparison of estimated stellar mass loss rates, $\dot{M_{*}}$ (calculated using Equation 2.1 in \citealt{Pellegrini12}, assuming a Kroupa IMF), with our calculated cooling flow rates, $\dot{M}$ (from the eighth column of \hyperref[Table4]{Table 4}). Galactic stellar masses, $M_*$, are quoted from \citealt{KormendyHo13}, \citealt{ZhuHoGao21}, \citealt{Iodice2019} and \citealt{Mathews21}. Ages of the galaxies in this sample were estimated to be about 10 Gyr \citep{Bregman06}.}
 \label{newtable}
\end{table}

{\renewcommand{\arraystretch}{1.1}
\begin{table}
\centering
 \begin{tblr}{c c c} 
 Target & ACF Luminosity & Estimated FIR Luminosity \\ 
 \hline
  & $10^8 \ L_\odot{}$ & $10^8 \ L_\odot{}$ \\
 \hline
 NGC 1316 & $0.91^{+0.5}_{-0.3}$ & $27.43^{+2.3}_{-5}$\\
 NGC 1332 & $0.74^{+0.2}_{-0.5}$ & $8.88^{+1.3}_{-1.2}$\\
 NGC 1404 & $1.08^{+0.4}_{-0.5}$ & $0.68^{+0.3}_{-0.4}$\\
 NGC 4552 & $0.08^{+0.04}_{-0.03}$ & $2.26^{+0.5}_{-0.4}$\\
 NGC 4636 & $0.09^{+0.05}_{-0.04}$ & $1.10^{+0.4}_{-0.3}$\\
 NGC 4649 & $0.85^{+0.2}_{-0.6}$ & $2.87^{+0.3}_{-0.6}$\\
 IC 1459 & $0.40^{+0.3}_{-0.4}$ & $9.16^{+1.4}_{-2.3}$\\
 \hline
 \end{tblr}
 \caption{Absorbed Cooling Flow (ACF) Luminosities calculated from HCF mass cooling flow rates in each object, compared to corresponding estimated FIR luminosities. Errors in ACF Luminosity are stated at the 68\% confidence interval according to the best-fit model. The FIR luminosities and associated errors were estimated using \citealt{Temi2007}, \citealt{Galametz2012} \& \citealt{Asabere2016}, along with data from the NASA Extragalactic Database.}
 \label{Table6}
\end{table}
}

\begin{figure*}
\centering{}
\includegraphics[width=\textwidth]{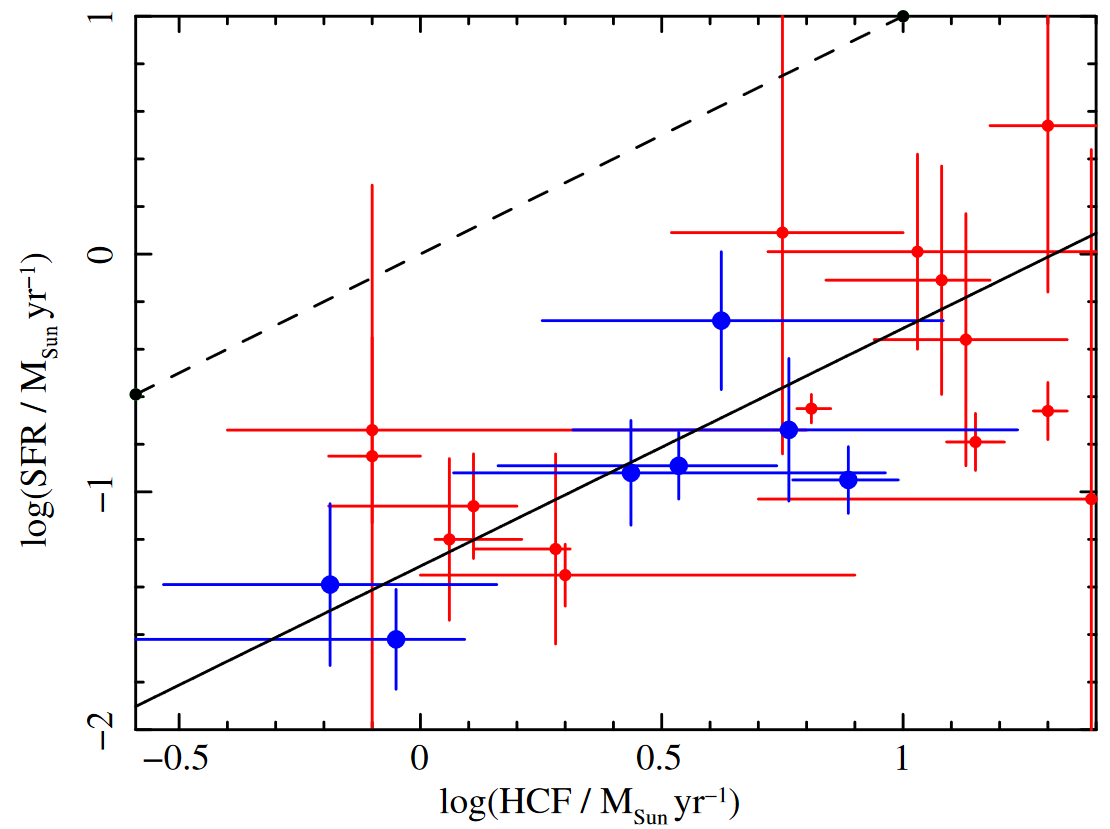}
    \caption{The mass cooling rate of the Hidden Cooling Flow plotted versus Star Formation Rate (SFR) in a sample of galaxies and galaxy clusters. SFRs and HCF rates are specified in units of $\textup{M}_\odot$ yr\textsuperscript{-1}. The red points are sourced from \citealt{HCF1, HCF2, HCF3, HCFlow4, Fabian2024}. SFRs for the red points were obtained from \citealt{McDonald2018}. The blue points are the sample of galaxies studied in this paper. The HCF rates are as presented in \hyperref[Table4]{Table 4}, and SFRs were determined using \citealt{Amblard2014} \& \citealt{Terrazas2017}. Error bars are set at the 90\% confidence interval. The upper dashed line is SFR = HCF. The lower solid line is SFR = HCF/20.}
    \label{Figure13}
\end{figure*}
\begin{figure}
\centering{}
\includegraphics[width=0.325\textwidth]{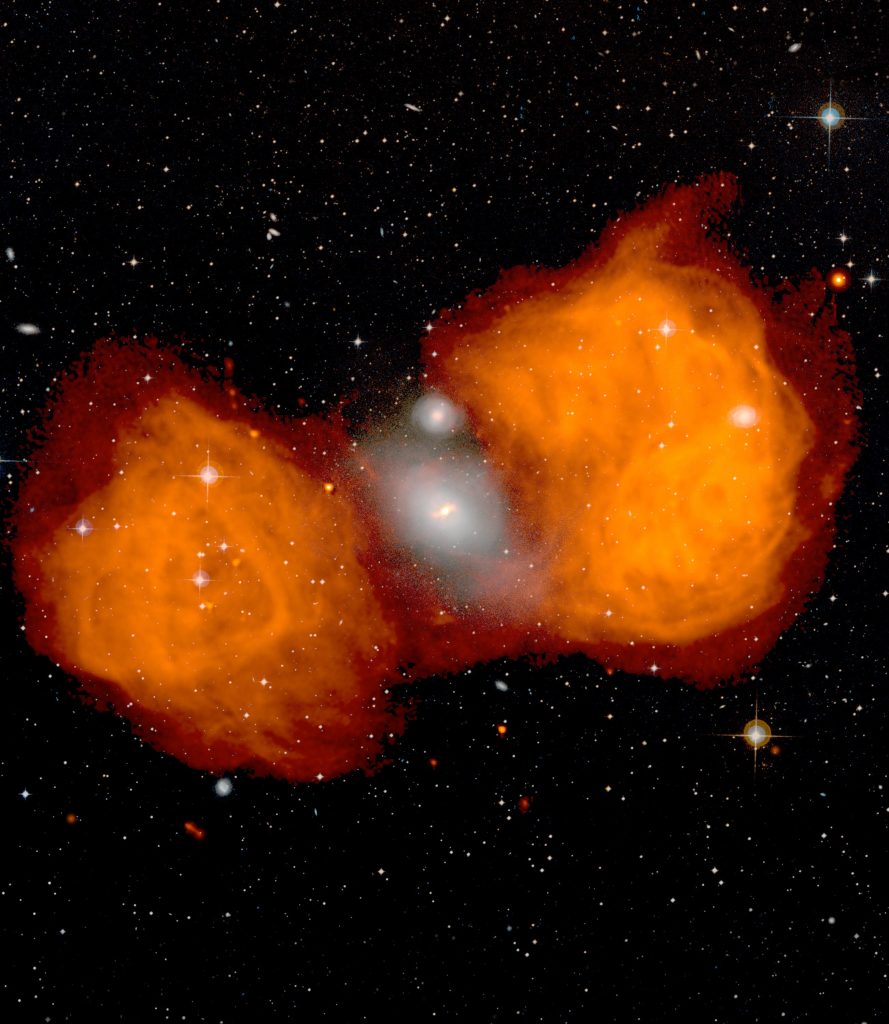}
    \caption{High resolution composite image of NGC 1316 by Juan Uson, combining the UK-Schmidt optical image with a 1.4 GHz VLA image; field of view is 55.5 $\times$ 63.9 arcmin (about 300 $\times $ 360 kpc), showing the galaxy's two radio lobes. \textit{Credit: NRAO/AUI/NSF; observers Fomalont, Ekers, van Breugel and Ebneter.}}
    \label{Figure 14}
\end{figure}

In this work, it has been shown that significant mass cooling rates are occurring in nearby elliptical galaxies, provided that gas cooling below 1 keV is interspersed with cold absorbing clouds with total column densities $\sim$$10^{22}$ cm\textsuperscript{-2}. This leads to the accumulation of much larger masses of cold gas than observed or generally considered. 

It is possible to compare our predicted cooling flow rates to the rates at which gas is returned to the ISM by stellar mass loss. The rate of stellar mass loss in each galaxy in our sample is shown in the third column of \hyperref[newtable]{Table 6}. For the majority of galaxies in our sample, stellar gas injection rates are significantly smaller than projected cooling rates. However, this does not represent a significant discrepancy - because the cooling flows are extremely localised \citep{HCFlow4}, extensive mixing of the cooling gas and the gas from stellar mass loss is unlikely. Furthermore, beyond the galactic centres there exist reservoirs of more than $10^{11} \ \text{M}_\odot$ of gas \citep{Lakhchaura18, Babyk2019}, which are more than sufficient to provide the few solar masses of cooling gas per year. While other heating processes can be at play, in this work we have focused on calculating the amount of gas which has an emission measure distribution that is compatible with a cooling flow.

It now remains to consider how feasible the calculated cooling flow rates are. If multilayer absorption allows significant mass cooling to occur below 1 keV, and the associated emission is emerging in the FIR, then a key question to address is how large the hidden cooling rate can be. Energy conservation is necessary, so total emission from longer wavelengths can provide a constraint. This is done by estimating the \textit{`Absorbed Cooling Flow Luminosity'}, the amount of soft X-ray luminosity from the cooling flow that is absorbed and then emitted at longer wavelengths. The Absorbed Cooling Flow Luminosity is calculated from intrinsic absorption model fits in XSPEC using \textsc{lumin}, by subtracting the luminosity calculated from the absorbed model to the luminosity calculated from a model with intrinsic column density set to zero. To set an upper bound on the Absorbed Cooling Flow Luminosity, the energy range considered was 0.1--3 keV, i.e. soft X-rays. The calculated Absorbed Cooling Flow Luminosities and associated errors are listed in the second column of \hyperref[Table6]{Table 7}.

The third column of \hyperref[Table6]{Table 7} presents FIR luminosity values from the literature. It can thus be seen that the absorbed luminosity in cooling flow is generally less than the FIR luminosity. The only exception to this is NGC 1404, but with errors taken into account there need not be a major discrepancy. While FIR luminosities alone are not definitive (SFR also contributes to FIR; see \cite{Kennicutt1998} for quantification of the relationship between SFR and FIR luminosity), these results indicate that energy lost to absorption in the cooling flow is energetically capable of emerging as long-wavelength radiation from dust in the absorbing gas.

The mass cooling rates we have found imply the accumulation of large cold masses, of the order of $10^9$--$10^{10} \ \textup{M}_\odot$, if they continue for several Gyr. Star formation is a possible consequence of this accumulation of mass.

\hyperref[Figure13]{Figure 13} shows a plot of star formation rates (SFRs) against predicted HCF rates in a sample of objects including those studied in this paper; the red points refer to data from \cite{Fabian2024}, while the blue points are data from galaxies studied in this paper. If SFRs and HCFs were one-to-one, we would expect the data points to be clustered around the upper dashed line. However, the points are instead clustered along the solid line, which indicates the observed SFRs are 20 times lower than what we would naively predict if all of the HCF mass went into star formation, similar to the ratio seen in clusters \citep{McDonald2018}. While star formation is not 100\% efficient, this is a massive discrepancy and does force us to consider what else could be happening to the accumulated cooled gas.

As discussed in \cite{HCF1, HCF2, HCF3}, several possible outcomes have been considered as the potential fate of cooled gas from HCFs, including AGN feedback reheating cooled gas and dragging it further out, or low-mass star formation. 

AGN feedback will not be discussed in detail here. As an example, \hyperref[Figure 14]{Figure 14} shows the radio lobes of NGC 1316, which are powered by the central AGN and extend $\sim$100 kpc from the centre of the galaxy. This indicates that energy from the AGN is redistributed many kpc away from the central region, not in the innermost $\sim$1 kpc where cooled material from an HCF accumulates. While NGC 1316 is an extreme object, this observation suggests that in galaxies with AGN, the impact of jetted feedback will be much more significant in the outer regions of the galaxy than in the central region. Hence, AGN feedback would be insufficient to explain reheating of cooled material near the galactic centre. 

AGN feedback is not necessarily jetted, and a more spherically symmetric mode of feedback could effectively deposit energy and momentum into the galactic centre. But one of the galaxies studied here (NGC 1404) does not show AGN activity at all, and still demonstrates the same Cooling Flow Problem as the others in our sample, so even if AGN feedback does play a role it cannot completely explain what we observe. Additionally, the Eddington ratio of most elliptical galaxies is highly sub-Eddington, which implies that the accretion flow in the centre is a low-luminosity ADAF, which we would not expect to produce radiation powerful enough to re-energise a significant amount of cooling gas; for further discussion of AGN feedback modes in elliptical galaxies see \cite{Ciotta09}.

It has been suggested that a `time average' of intermittent cooling and heating from AGN feedback could match our observations. However, there is no evidence of sporadic star formation in the centres of massive elliptical galaxies. We are also assuming that cooling flows are continuous and steady - we will discuss later in this section a new probe that we hope can, in the future, be used to demonstrate that this is the case.

Overall, to match observations, AGN feedback would need to be in close balance with any cooling that takes place. This presents a fine tuning problem, which indicates that AGN feedback is not a feasible solution to the Cooling Flow Problem at the centres of massive elliptical galaxies. 

Instead, we propose low-mass star formation is the primary outcome for the cooled material in HCFs. The centres of elliptical galaxies are high pressure environments, which means they have a low Jeans mass and so low-mass star formation is favoured. Evidence for this process comes from evidence for bottom heavy IMFs near the centres of early-type galaxies, as found by \cite{vanDokkum2010} and later reviewed more generally by \cite{Smith2020}. Recently, \cite{Gu22} have determined the stellar IMF in the centres  of 41 nearby massive early-type galaxies, requiring a steep low-mass IMF slope in all. NGC 4649, which we have discussed in this paper, is also included in their sample.

It was shown in \cite{HCF1} that under the high pressure conditions of an HCF with no heating, the gas cools rapidly to $\sim$3 K. The Jeans mass, defined by 
\begin{equation}
    M_\text{J} = \frac{\pi}{6} \frac{c_s^3}{G^{3/2} \rho^{1/2}},
\end{equation}
is the critical mass at which gas clouds becomes unstable to gravitational collapse, causing them to fragment ($c_s$ refers to the speed of sound in the gas, and $\rho$ to its density). If the Jeans mass is sufficiently large, the collapsed gas clouds are massive enough to start hydrogen fusion and hence form stars upon their collapse; however, if the gas is very cold and dense then the Jeans mass can be much lower than this. Consequently, the collapsed gas never becomes hot enough to commence hydrogen fusion, and low-mass objects such as brown dwarfs (which are powered by deuterium fusion) are formed instead. In the high-pressure centres of galaxies, the Jeans mass is below about 0.1 M\textsubscript{$\odot$} \citep{Jura1977, Ferland1994} and a low-mass star formation mode may operate there \citep{Fabian2024}. 

\cite{2024vandokkumconroy} discuss and propose a concordance IMF for massive galaxies. It has a `bottom heavy' part below 0.5 M\textsubscript{$\odot$}, with a steep IMF slope of 2.4 for the innermost region. We propose that this steep part is produced by the cooling flow, as described above. 

Over long timescales, low mass star formation can result in the accumulation of large, `hidden' masses in the galactic centre, which we would expect to have a significant impact on the stellar dynamics there. Furthermore, based on spectral data alone, we do not know the spatial distribution of the cooling flows or the resulting low mass star formation. It is entirely possible that the remaining angular momentum of the cooling gas is causing the resulting 'hidden' mass to settle into a disk, for example the dark disk-shape that we can see within NGC 1332 (see \hyperref[11b]{Figure 11(b)}). However, we will not speculate much further on these matters, as we hope future studies of these objects will be able to unveil the exact structure of the galactic centres and tell us more about how cooling flows affect stellar kinematics.

The formation of low-mass stars can result in `hidden' accretion onto the central black holes of galaxies. The accretion is hidden because low-mass stars may be swallowed whole without emitting radiation. As a result it is possible that not only the mass of the dark cores of elliptical galaxies but also the mass of black holes in nearby elliptical galaxies is increasing due to hidden cooling flows (see \citealt{HCF3} and \citealt{2023farrah}).

An interesting new probe of the nature of absorbed cooling flows has recently been proposed \citep{HCFlow4}. The [Ne VI] emission line at $7.652 \ \mu\text{m}$ in the mid-infrared is indicative of gas cooling through temperatures of $6-1.5 \times 10^5$ K, and based on our preliminary calculations this emission line should be detectable in some nearby massive elliptical galaxies using JWST's MIRI-IFU. We can use measurements of this emission line to show that gas in the centres of elliptical galaxies is in fact cooling below $10^6$ K, demonstrating that absorbed cooling flows exist in these objects. Measurements of [Ne VI] flux can constrain the mass cooling rates in the galactic centres, and increase confidence in the HCF rates predicted by the Intrinsic Absorption Model. This emission line can also give us information about the kinematics of the cooling flow, providing insight into how the cooling flows operate (i.e. whether they are continuous and steady) and the ways in which they impact the dynamics of the galactic centre.

\section{Conclusions}
Significant hidden cooling flows are common in nearby elliptical galaxies, with mass cooling rates of $\sim$0.5--8$\ \text{M}_{\odot}$ yr\textsuperscript{-1}. Evidence for cold gas near the galactic centres has been found via HST images of absorption in the galaxies studied. Intrinsic photoelectric absorption explains the observed lack of strong soft X-ray emission from cooling flows in the spectra of elliptical galaxies; the emission instead emerges from the absorbing material at long wavelengths, including the far infrared. Cooled gas flowing into the central, absorbed part of the galaxy can cool below 10 K, collapsing and fragmenting into low-mass stars and brown dwarfs. Some of these objects may fall into the galaxy's central black hole, emitting very little radiation. The centres of massive elliptical galaxies may be quiescent in terms of {\it normal} star formation but they are quite the opposite in low-mass star formation.

Further multi-wavelength studies of the centres of elliptical galaxies will be key to understanding the environment and nature of cooling flows within them. For example, JWST, which operates in the near- to mid-infrared, will enable mapping and quantification of the cold components of the galaxies; in particular, the mid-infrared [Ne VI] emission line is a promising probe of the extent and nature of absorbed cooling flows in these objects. The proposed X-ray probe AXIS with CCD spectral resolution and 1 arcsec spatial resolution, and the XIFU of NewAthena with 5 arcsec pixels and 4 eV spectral resolution, will map galaxy centres in higher detail, enabling spatial distribution of mass cooling rate and column density to be correlated with molecular gas and dust. This will reveal in depth how the cooling flow operates. The objects studied in this paper represent excellent potential targets for both of these missions.

Ultimately, in this work we have demonstrated the success of the intrinsic absorption model in fitting spectra of nearby elliptical galaxies. It represents a further step towards a resolution of the Cooling Flow Problem. 

\section*{Acknowledgements}
Many thanks to Annabelle Richard-Laferriere for help with software, and to Roberto Maiolino for funding the completion of this research. Thanks also to the referee, Professor Luca Ciotti, for his insightful comments.

%%%%%%%%%%%%%%%%%%%%%%%%%%%%%%%%%%%%%%%%%%%%%%%%%%
\section*{Data Availability}
All data used here are available from ESA’s XMM-Newton Science
Archive, the Hubble Legacy Archive, and NASA’s Chandra Data Archive.



\bibliographystyle{mnras}
\bibliography{example} % if your bibtex file is called example.bib

\begin{thebibliography}{}
\makeatletter
\relax
\def\mn@urlcharsother{\let\do\@makeother \do\$\do\&\do\#\do\^\do\_\do\%\do\~}
\def\mn@doi{\begingroup\mn@urlcharsother \@ifnextchar [ {\mn@doi@} {\mn@doi@[]}}
\def\mn@doi@[#1]#2{\def\@tempa{#1}\ifx\@tempa\@empty \href {http://dx.doi.org/#2} {doi:#2}\else \href {http://dx.doi.org/#2} {#1}\fi \endgroup}
\def\mn@eprint#1#2{\mn@eprint@#1:#2::\@nil}
\def\mn@eprint@arXiv#1{\href {http://arxiv.org/abs/#1} {{\tt arXiv:#1}}}
\def\mn@eprint@dblp#1{\href {http://dblp.uni-trier.de/rec/bibtex/#1.xml} {dblp:#1}}
\def\mn@eprint@#1:#2:#3:#4\@nil{\def\@tempa {#1}\def\@tempb {#2}\def\@tempc {#3}\ifx \@tempc \@empty \let \@tempc \@tempb \let \@tempb \@tempa \fi \ifx \@tempb \@empty \def\@tempb {arXiv}\fi \@ifundefined {mn@eprint@\@tempb}{\@tempb:\@tempc}{\expandafter \expandafter \csname mn@eprint@\@tempb\endcsname \expandafter{\@tempc}}}

\bibitem[\protect\citeauthoryear{{Allen}}{{Allen}}{2000}]{Allen2000}
{Allen} S.~W.,  2000, \mn@doi [\mnras] {10.1046/j.1365-8711.2000.03395.x}, \href {https://ui.adsabs.harvard.edu/abs/2000MNRAS.315..269A} {315, 269}

\bibitem[\protect\citeauthoryear{{Allen} \& {Fabian}}{{Allen} \& {Fabian}}{1997}]{Allen&Fabian1997spatial}
{Allen} S.~W.,  {Fabian} A.~C.,  1997, \mn@doi [\mnras] {10.1093/mnras/286.3.583}, \href {https://ui.adsabs.harvard.edu/abs/1997MNRAS.286..583A} {286, 583}

\bibitem[\protect\citeauthoryear{{Allen}, {Fabian}, {Johnstone}, {Arnaud}  \& {Nulsen}}{{Allen} et~al.}{2001}]{AllenEtAL2001}
{Allen} S.~W.,  {Fabian} A.~C.,  {Johnstone} R.~M.,  {Arnaud} K.~A.,   {Nulsen} P.~E.~J.,  2001, \mn@doi [\mnras] {10.1046/j.1365-8711.2001.04135.x}, \href {https://ui.adsabs.harvard.edu/abs/2001MNRAS.322..589A} {322, 589}

\bibitem[\protect\citeauthoryear{{Allen}, {Dunn}, {Fabian}, {Taylor}  \& {Reynolds}}{{Allen} et~al.}{2006}]{Allen2006}
{Allen} S.~W.,  {Dunn} R.~J.~H.,  {Fabian} A.~C.,  {Taylor} G.~B.,   {Reynolds} C.~S.,  2006, \mn@doi [\mnras] {10.1111/j.1365-2966.2006.10778.x}, \href {https://ui.adsabs.harvard.edu/abs/2006MNRAS.372...21A} {372, 21}

\bibitem[\protect\citeauthoryear{{Amblard}, {Riguccini}, {Temi}, {Im}, {Fanelli}  \& {Serra}}{{Amblard} et~al.}{2014}]{Amblard2014}
{Amblard} A.,  {Riguccini} L.,  {Temi} P.,  {Im} S.,  {Fanelli} M.,   {Serra} P.,  2014, \mn@doi [\apj] {10.1088/0004-637X/783/2/135}, \href {https://ui.adsabs.harvard.edu/abs/2014ApJ...783..135A} {783, 135}

\bibitem[\protect\citeauthoryear{{Arnaud}}{{Arnaud}}{1996}]{Arnaud1996}
{Arnaud} K.~A.,  1996, in {Jacoby} G.~H.,  {Barnes} J.,  eds,  Astronomical Society of the Pacific Conference Series Vol. 101, Astronomical Data Analysis Software and Systems V. p.~17

\bibitem[\protect\citeauthoryear{{Babyk}, {McNamara}, {Tamhane}, {Nulsen}, {Russell}  \& {Edge}}{{Babyk} et~al.}{2019}]{Babyk2019}
{Babyk} I.~V.,  {McNamara} B.~R.,  {Tamhane} P.~D.,  {Nulsen} P.~E.~J.,  {Russell} H.~R.,   {Edge} A.~C.,  2019, \mn@doi [APJ] {10.3847/1538-4357/ab54ce}, \href {https://ui.adsabs.harvard.edu/abs/2019ApJ...887..149B} {887, 149}

\bibitem[\protect\citeauthoryear{{Baldi}, {Forman}, {Jones}, {Kraft}, {Nulsen}, {Churazov}, {David}  \& {Giacintucci}}{{Baldi} et~al.}{2009}]{Baldi2009}
{Baldi} A.,  {Forman} W.,  {Jones} C.,  {Kraft} R.,  {Nulsen} P.,  {Churazov} E.,  {David} L.,   {Giacintucci} S.,  2009, \mn@doi [\apj] {10.1088/0004-637X/707/2/1034}, \href {https://ui.adsabs.harvard.edu/abs/2009ApJ...707.1034B} {707, 1034}

\bibitem[\protect\citeauthoryear{{Barth}, {Boizelle}, {Darling}, {Baker}, {Buote}, {Ho}  \& {Walsh}}{{Barth} et~al.}{2016}]{Barth2016}
{Barth} A.~J.,  {Boizelle} B.,  {Darling} J.~K.,  {Baker} A.~J.,  {Buote} D.~A.,  {Ho} L.~C.,   {Walsh} J.,  2016, in American Astronomical Society Meeting Abstracts \#228. p. 103.05

\bibitem[\protect\citeauthoryear{{Bekki}, {Forbes}, {Beasley}  \& {Couch}}{{Bekki} et~al.}{2003}]{Bekki2003}
{Bekki} K.,  {Forbes} D.~A.,  {Beasley} M.~A.,   {Couch} W.~J.,  2003, \mn@doi [\mnras] {10.1046/j.1365-8711.2003.06925.x}, \href {https://ui.adsabs.harvard.edu/abs/2003MNRAS.344.1334B} {344, 1334}

\bibitem[\protect\citeauthoryear{{Bregman}, {Miller}  \& {Irwin}}{{Bregman} et~al.}{2001}]{Bregman2001}
{Bregman} J.~N.,  {Miller} E.~D.,   {Irwin} J.~A.,  2001, \mn@doi [\apjl] {10.1086/320675}, \href {https://ui.adsabs.harvard.edu/abs/2001ApJ...553L.125B} {553, L125}

\bibitem[\protect\citeauthoryear{{Bregman}, {Temi}  \& {Bregman}}{{Bregman} et~al.}{2006}]{Bregman06}
{Bregman} J.~N.,  {Temi} P.,   {Bregman} J.~D.,  2006, \mn@doi [\apj] {10.1086/505190}, \href {https://ui.adsabs.harvard.edu/abs/2006ApJ...647..265B} {647, 265}

\bibitem[\protect\citeauthoryear{{Br{\"u}ggen}, {Ruszkowski}  \& {Hallman}}{{Br{\"u}ggen} et~al.}{2005}]{2005Bruggen}
{Br{\"u}ggen} M.,  {Ruszkowski} M.,   {Hallman} E.,  2005, \mn@doi [\apj] {10.1086/432112}, \href {https://ui.adsabs.harvard.edu/abs/2005ApJ...630..740B} {630, 740}

\bibitem[\protect\citeauthoryear{{Cappellari}, {Verolme}, {van der Marel}, {Verdoes Kleijn}, {Illingworth}, {Franx}, {Carollo}  \& {de Zeeuw}}{{Cappellari} et~al.}{2002}]{Cappellari2002}
{Cappellari} M.,  {Verolme} E.~K.,  {van der Marel} R.~P.,  {Verdoes Kleijn} G.~A.,  {Illingworth} G.~D.,  {Franx} M.,  {Carollo} C.~M.,   {de Zeeuw} P.~T.,  2002, \mn@doi [\apj] {10.1086/342653}, \href {https://ui.adsabs.harvard.edu/abs/2002ApJ...578..787C} {578, 787}

\bibitem[\protect\citeauthoryear{{Ciotti}, {Ostriker}  \& {Proga}}{{Ciotti} et~al.}{2009}]{Ciotta09}
{Ciotti} L.,  {Ostriker} J.~P.,   {Proga} D.,  2009, \mn@doi [\apj] {10.1088/0004-637X/699/1/89}, \href {https://ui.adsabs.harvard.edu/abs/2009ApJ...699...89C} {699, 89}

\bibitem[\protect\citeauthoryear{{Cowie} \& {Binney}}{{Cowie} \& {Binney}}{1977}]{CowieBinney1977}
{Cowie} L.~L.,  {Binney} J.,  1977, \mn@doi [\apj] {10.1086/155406}, \href {https://ui.adsabs.harvard.edu/abs/1977ApJ...215..723C} {215, 723}

\bibitem[\protect\citeauthoryear{{Crawford} \& {Fabian}}{{Crawford} \& {Fabian}}{1992}]{1992CrawfordFabian}
{Crawford} C.~S.,  {Fabian} A.~C.,  1992, \mn@doi [\mnras] {10.1093/mnras/259.2.265}, \href {https://ui.adsabs.harvard.edu/abs/1992MNRAS.259..265C} {259, 265}

\bibitem[\protect\citeauthoryear{{Dickey} \& {Lockman}}{{Dickey} \& {Lockman}}{1990}]{Dickey1990}
{Dickey} J.~M.,  {Lockman} F.~J.,  1990, \mn@doi [\araa] {10.1146/annurev.aa.28.090190.001243}, \href {https://ui.adsabs.harvard.edu/abs/1990ARA&A..28..215D} {28, 215}

\bibitem[\protect\citeauthoryear{{Draine} \& {Salpeter}}{{Draine} \& {Salpeter}}{1979}]{Draine1979}
{Draine} B.~T.,  {Salpeter} E.~E.,  1979, \mn@doi [\apj] {10.1086/157165}, \href {https://ui.adsabs.harvard.edu/abs/1979ApJ...231...77D} {231, 77}

\bibitem[\protect\citeauthoryear{{Duah Asabere}, {Horellou}, {Jarrett}  \& {Winkler}}{{Duah Asabere} et~al.}{2016}]{Asabere2016}
{Duah Asabere} B.,  {Horellou} C.,  {Jarrett} T.~H.,   {Winkler} H.,  2016, \mn@doi [\aap] {10.1051/0004-6361/201528047}, \href {https://ui.adsabs.harvard.edu/abs/2016A&A...592A..20D} {592, A20}

\bibitem[\protect\citeauthoryear{{Fabbiano} et~al.,}{{Fabbiano} et~al.}{2003}]{Fabbiano2003}
{Fabbiano} G.,  et~al., 2003, \mn@doi [\apj] {10.1086/374040}, \href {https://ui.adsabs.harvard.edu/abs/2003ApJ...588..175F} {588, 175}

\bibitem[\protect\citeauthoryear{{Fabian}}{{Fabian}}{1994}]{Fabian1994}
{Fabian} A.~C.,  1994, \mn@doi [Annual Review of Astronomy and Astrophysics] {10.1146/annurev.aa.32.090194.001425}, \href {https://ui.adsabs.harvard.edu/abs/1994ARA&A..32..277F} {32, 277}

\bibitem[\protect\citeauthoryear{{Fabian} \& {Nulsen}}{{Fabian} \& {Nulsen}}{1977}]{fabiannulsen1977}
{Fabian} A.~C.,  {Nulsen} P.~E.~J.,  1977, \mn@doi [\mnras] {10.1093/mnras/180.3.479}, \href {https://ui.adsabs.harvard.edu/abs/1977MNRAS.180..479F} {180, 479}

\bibitem[\protect\citeauthoryear{{Fabian}, {Arnaud}, {Bautz}  \& {Tawara}}{{Fabian} et~al.}{1994}]{Fabian1994abs}
{Fabian} A.~C.,  {Arnaud} K.~A.,  {Bautz} M.~W.,   {Tawara} Y.,  1994, \mn@doi [\apjl] {10.1086/187633}, \href {https://ui.adsabs.harvard.edu/abs/1994ApJ...436L..63F} {436, L63}

\bibitem[\protect\citeauthoryear{{Fabian}, {Ferland}, {Sanders}, {McNamara}, {Pinto}  \& {Walker}}{{Fabian} et~al.}{2022}]{HCF1}
{Fabian} A.~C.,  {Ferland} G.~J.,  {Sanders} J.~S.,  {McNamara} B.~R.,  {Pinto} C.,   {Walker} S.~A.,  2022, \mn@doi [MNRAS] {10.1093/mnras/stac2003}, \href {https://ui.adsabs.harvard.edu/abs/2022MNRAS.515.3336F} {515, 3336}

\bibitem[\protect\citeauthoryear{{Fabian}, {Sanders}, {Ferland}, {McNamara}, {Pinto}  \& {Walker}}{{Fabian} et~al.}{2023a}]{HCF2}
{Fabian} A.~C.,  {Sanders} J.~S.,  {Ferland} G.~J.,  {McNamara} B.~R.,  {Pinto} C.,   {Walker} S.~A.,  2023a, \mn@doi [MNRAS] {10.1093/mnras/stad507}, \href {https://ui.adsabs.harvard.edu/abs/2023MNRAS.521.1794F} {521, 1794}

\bibitem[\protect\citeauthoryear{{Fabian}, {Sanders}, {Ferland}, {McNamara}, {Pinto}  \& {Walker}}{{Fabian} et~al.}{2023b}]{HCF3}
{Fabian} A.~C.,  {Sanders} J.~S.,  {Ferland} G.~J.,  {McNamara} B.~R.,  {Pinto} C.,   {Walker} S.~A.,  2023b, \mn@doi [MNRAS] {10.1093/mnras/stad1870}, \href {https://ui.adsabs.harvard.edu/abs/2023MNRAS.524..716F} {524, 716}

\bibitem[\protect\citeauthoryear{{Fabian} et~al.,}{{Fabian} et~al.}{2024a}]{HCFlow4}
{Fabian} A.~C.,  et~al., 2024a, \mn@doi [arXiv e-prints] {10.48550/arXiv.2410.10675}, \href {https://ui.adsabs.harvard.edu/abs/2024arXiv241010675F} {p. arXiv:2410.10675}

\bibitem[\protect\citeauthoryear{{Fabian}, {Sanders}, {Ferland}, {McNamara}, {Pinto}  \& {Walker}}{{Fabian} et~al.}{2024b}]{Fabian2024}
{Fabian} A.~C.,  {Sanders} J.~S.,  {Ferland} G.~J.,  {McNamara} B.~R.,  {Pinto} C.,   {Walker} S.~A.,  2024b, \mn@doi [\mnras] {10.1093/mnras/stae1206}, \href {https://ui.adsabs.harvard.edu/abs/2024MNRAS.531..267F} {531, 267}

\bibitem[\protect\citeauthoryear{{Farrah} et~al.,}{{Farrah} et~al.}{2023}]{2023farrah}
{Farrah} D.,  et~al., 2023, \mn@doi [\apj] {10.3847/1538-4357/acac2e}, \href {https://ui.adsabs.harvard.edu/abs/2023ApJ...943..133F} {943, 133}

\bibitem[\protect\citeauthoryear{{Ferguson}}{{Ferguson}}{1989}]{Ferguson1989}
{Ferguson} H.~C.,  1989, \mn@doi [\aj] {10.1086/115152}, \href {https://ui.adsabs.harvard.edu/abs/1989AJ.....98..367F} {98, 367}

\bibitem[\protect\citeauthoryear{{Ferland}, {Fabian}  \& {Johnstone}}{{Ferland} et~al.}{1994}]{Ferland1994}
{Ferland} G.~J.,  {Fabian} A.~C.,   {Johnstone} R.~M.,  1994, \mn@doi [\mnras] {10.1093/mnras/266.2.399}, \href {https://ui.adsabs.harvard.edu/abs/1994MNRAS.266..399F} {266, 399}

\bibitem[\protect\citeauthoryear{{Forbes}, {Grillmair}, {Williger}, {Elson}  \& {Brodie}}{{Forbes} et~al.}{1998}]{Forbes1998}
{Forbes} D.~A.,  {Grillmair} C.~J.,  {Williger} G.~M.,  {Elson} R.~A.~W.,   {Brodie} J.~P.,  1998, \mn@doi [\mnras] {10.1046/j.1365-8711.1998.01202.x}, \href {https://ui.adsabs.harvard.edu/abs/1998MNRAS.293..325F} {293, 325}

\bibitem[\protect\citeauthoryear{{Forman}, {Jones}  \& {Tucker}}{{Forman} et~al.}{1985}]{Forman1985}
{Forman} W.,  {Jones} C.,   {Tucker} W.,  1985, \mn@doi [\apj] {10.1086/163218}, \href {https://ui.adsabs.harvard.edu/abs/1985ApJ...293..102F} {293, 102}

\bibitem[\protect\citeauthoryear{{Gabriel} et~al.,}{{Gabriel} et~al.}{2004}]{Gabriel2004}
{Gabriel} C.,  et~al., 2004, in {Ochsenbein} F.,  {Allen} M.~G.,   {Egret} D.,  eds,  Astronomical Society of the Pacific Conference Series Vol. 314, Astronomical Data Analysis Software and Systems (ADASS) XIII. p.~759

\bibitem[\protect\citeauthoryear{{Galametz} et~al.,}{{Galametz} et~al.}{2012}]{Galametz2012}
{Galametz} M.,  et~al., 2012, \mn@doi [\mnras] {10.1111/j.1365-2966.2012.21667.x}, \href {https://ui.adsabs.harvard.edu/abs/2012MNRAS.425..763G} {425, 763}

\bibitem[\protect\citeauthoryear{{Giacintucci} et~al.,}{{Giacintucci} et~al.}{2011}]{Giacintucci2011}
{Giacintucci} S.,  et~al., 2011, \mn@doi [\apj] {10.1088/0004-637X/732/2/95}, \href {https://ui.adsabs.harvard.edu/abs/2011ApJ...732...95G} {732, 95}

\bibitem[\protect\citeauthoryear{Graham}{Graham}{2008}]{Graham_2008}
Graham A.~W.,  2008, \mn@doi [Publications of the Astronomical Society of Australia] {10.1071/as08013}, 25, 167–175

\bibitem[\protect\citeauthoryear{{Gu}, {Greene}, {Newman}, {Kreisch}, {Quenneville}, {Ma}  \& {Blakeslee}}{{Gu} et~al.}{2022}]{Gu22}
{Gu} M.,  {Greene} J.~E.,  {Newman} A.~B.,  {Kreisch} C.,  {Quenneville} M.~E.,  {Ma} C.-P.,   {Blakeslee} J.~P.,  2022, \mn@doi [\apj] {10.3847/1538-4357/ac69ea}, \href {https://ui.adsabs.harvard.edu/abs/2022ApJ...932..103G} {932, 103}

\bibitem[\protect\citeauthoryear{{Ho}, {Filippenko}, {Sargent}  \& {Peng}}{{Ho} et~al.}{1997}]{Ho1997}
{Ho} L.~C.,  {Filippenko} A.~V.,  {Sargent} W. L.~W.,   {Peng} C.~Y.,  1997, \mn@doi [\apjs] {10.1086/313042}, \href {https://ui.adsabs.harvard.edu/abs/1997ApJS..112..391H} {112, 391}

\bibitem[\protect\citeauthoryear{{Humphrey} \& {Buote}}{{Humphrey} \& {Buote}}{2004}]{Humphrey1}
{Humphrey} P.~J.,  {Buote} D.~A.,  2004, \mn@doi [\apj] {10.1086/422743}, \href {https://ui.adsabs.harvard.edu/abs/2004ApJ...612..848H} {612, 848}

\bibitem[\protect\citeauthoryear{{Humphrey}, {Buote}  \& {Canizares}}{{Humphrey} et~al.}{2004}]{Humphrey2}
{Humphrey} P.~J.,  {Buote} D.~A.,   {Canizares} C.~R.,  2004, \mn@doi [\apj] {10.1086/425413}, \href {https://ui.adsabs.harvard.edu/abs/2004ApJ...617.1047H} {617, 1047}

\bibitem[\protect\citeauthoryear{{Iodice} et~al.,}{{Iodice} et~al.}{2019}]{Iodice2019}
{Iodice} E.,  et~al., 2019, \mn@doi [\aap] {10.1051/0004-6361/201833741}, \href {https://ui.adsabs.harvard.edu/abs/2019A&A...623A...1I} {623, A1}

\bibitem[\protect\citeauthoryear{{Irwin}, {Athey}  \& {Bregman}}{{Irwin} et~al.}{2003}]{Irwin2003}
{Irwin} J.~A.,  {Athey} A.~E.,   {Bregman} J.~N.,  2003, \mn@doi [\apj] {10.1086/368179}, \href {https://ui.adsabs.harvard.edu/abs/2003ApJ...587..356I} {587, 356}

\bibitem[\protect\citeauthoryear{Janowiecki, Mihos, Harding, Feldmeier, Rudick  \& Morrison}{Janowiecki et~al.}{2010}]{Janowiecki_2010}
Janowiecki S.,  Mihos J.~C.,  Harding P.,  Feldmeier J.~J.,  Rudick C.,   Morrison H.,  2010, \mn@doi [The Astrophysical Journal] {10.1088/0004-637x/715/2/972}, 715, 972–985

\bibitem[\protect\citeauthoryear{{Johnstone}, {Fabian}  \& {Nulsen}}{{Johnstone} et~al.}{1987}]{1987JFN}
{Johnstone} R.~M.,  {Fabian} A.~C.,   {Nulsen} P.~E.~J.,  1987, \mn@doi [\mnras] {10.1093/mnras/224.1.75}, \href {https://ui.adsabs.harvard.edu/abs/1987MNRAS.224...75J} {224, 75}

\bibitem[\protect\citeauthoryear{{Johnstone}, {Fabian}, {Edge}  \& {Thomas}}{{Johnstone} et~al.}{1992}]{JohnstoneEtAl1992}
{Johnstone} R.~M.,  {Fabian} A.~C.,  {Edge} A.~C.,   {Thomas} P.~A.,  1992, \mn@doi [\mnras] {10.1093/mnras/255.3.431}, \href {https://ui.adsabs.harvard.edu/abs/1992MNRAS.255..431J} {255, 431}

\bibitem[\protect\citeauthoryear{{Jura}}{{Jura}}{1977}]{Jura1977}
{Jura} M.,  1977, \mn@doi [\apj] {10.1086/155085}, \href {https://ui.adsabs.harvard.edu/abs/1977ApJ...212..634J} {212, 634}

\bibitem[\protect\citeauthoryear{{Kalberla}, {Burton}, {Hartmann}, {Arnal}, {Bajaja}, {Morras}  \& {P{\"o}ppel}}{{Kalberla} et~al.}{2005}]{Kalberla2005}
{Kalberla} P.~M.~W.,  {Burton} W.~B.,  {Hartmann} D.,  {Arnal} E.~M.,  {Bajaja} E.,  {Morras} R.,   {P{\"o}ppel} W.~G.~L.,  2005, \mn@doi [\aap] {10.1051/0004-6361:20041864}, \href {https://ui.adsabs.harvard.edu/abs/2005A&A...440..775K} {440, 775}

\bibitem[\protect\citeauthoryear{{Kennicutt}}{{Kennicutt}}{1998}]{Kennicutt1998}
{Kennicutt} Robert~C. J.,  1998, \mn@doi [\apj] {10.1086/305588}, \href {https://ui.adsabs.harvard.edu/abs/1998ApJ...498..541K} {498, 541}

\bibitem[\protect\citeauthoryear{{Kim} \& {Fabbiano}}{{Kim} \& {Fabbiano}}{2003}]{KimFabbiano2003}
{Kim} D.-W.,  {Fabbiano} G.,  2003, \mn@doi [\apj] {10.1086/367930}, \href {https://ui.adsabs.harvard.edu/abs/2003ApJ...586..826K} {586, 826}

\bibitem[\protect\citeauthoryear{{Kim} \& {Fabbiano}}{{Kim} \& {Fabbiano}}{2004}]{KimFabbiano2004}
{Kim} D.-W.,  {Fabbiano} G.,  2004, \mn@doi [\apj] {10.1086/422210}, \href {https://ui.adsabs.harvard.edu/abs/2004ApJ...611..846K} {611, 846}

\bibitem[\protect\citeauthoryear{{Kormendy} \& {Ho}}{{Kormendy} \& {Ho}}{2013}]{KormendyHo13}
{Kormendy} J.,  {Ho} L.~C.,  2013, \mn@doi [\araa] {10.1146/annurev-astro-082708-101811}, \href {https://ui.adsabs.harvard.edu/abs/2013ARA&A..51..511K} {51, 511}

\bibitem[\protect\citeauthoryear{{Lakhchaura} et~al.,}{{Lakhchaura} et~al.}{2018}]{Lakhchaura18}
{Lakhchaura} K.,  et~al., 2018, \mn@doi [\mnras] {10.1093/mnras/sty2565}, \href {https://ui.adsabs.harvard.edu/abs/2018MNRAS.481.4472L} {481, 4472}

\bibitem[\protect\citeauthoryear{{Liu}, {Fabian}, {Pinto}, {Russell}, {Sanders}  \& {McNamara}}{{Liu} et~al.}{2021}]{Liu2021}
{Liu} H.,  {Fabian} A.~C.,  {Pinto} C.,  {Russell} H.~R.,  {Sanders} J.~S.,   {McNamara} B.~R.,  2021, \mn@doi [\mnras] {10.1093/mnras/stab1372}, \href {https://ui.adsabs.harvard.edu/abs/2021MNRAS.505.1589L} {505, 1589}

\bibitem[\protect\citeauthoryear{{Machacek}, {Dosaj}, {Forman}, {Jones}, {Markevitch}, {Vikhlinin}, {Warmflash}  \& {Kraft}}{{Machacek} et~al.}{2005}]{Machacek2005}
{Machacek} M.,  {Dosaj} A.,  {Forman} W.,  {Jones} C.,  {Markevitch} M.,  {Vikhlinin} A.,  {Warmflash} A.,   {Kraft} R.,  2005, \mn@doi [\apj] {10.1086/427548}, \href {https://ui.adsabs.harvard.edu/abs/2005ApJ...621..663M} {621, 663}

\bibitem[\protect\citeauthoryear{{Machacek}, {Jones}, {Forman}  \& {Nulsen}}{{Machacek} et~al.}{2006a}]{Machacek2006strip}
{Machacek} M.,  {Jones} C.,  {Forman} W.~R.,   {Nulsen} P.,  2006a, \mn@doi [\apj] {10.1086/503350}, \href {https://ui.adsabs.harvard.edu/abs/2006ApJ...644..155M} {644, 155}

\bibitem[\protect\citeauthoryear{{Machacek}, {Nulsen}, {Jones}  \& {Forman}}{{Machacek} et~al.}{2006b}]{Machacek2006}
{Machacek} M.,  {Nulsen} P.~E.~J.,  {Jones} C.,   {Forman} W.~R.,  2006b, \mn@doi [\apj] {10.1086/505963}, \href {https://ui.adsabs.harvard.edu/abs/2006ApJ...648..947M} {648, 947}

\bibitem[\protect\citeauthoryear{{Mathews}}{{Mathews}}{2021}]{Mathews21}
{Mathews} W.~G.,  2021, \mn@doi [\mnras] {10.1093/mnras/stab1745}, \href {https://ui.adsabs.harvard.edu/abs/2021MNRAS.506.2030M} {506, 2030}

\bibitem[\protect\citeauthoryear{{McDonald}, {Gaspari}, {McNamara}  \& {Tremblay}}{{McDonald} et~al.}{2018}]{McDonald2018}
{McDonald} M.,  {Gaspari} M.,  {McNamara} B.~R.,   {Tremblay} G.~R.,  2018, \mn@doi [\apj] {10.3847/1538-4357/aabace}, \href {https://ui.adsabs.harvard.edu/abs/2018ApJ...858...45M} {858, 45}

\bibitem[\protect\citeauthoryear{Mei et~al.,}{Mei et~al.}{2007}]{Mei_2007}
Mei S.,  et~al., 2007, \mn@doi [The Astrophysical Journal] {10.1086/509598}, 655, 144–162

\bibitem[\protect\citeauthoryear{{Merritt} \& {Ferrarese}}{{Merritt} \& {Ferrarese}}{2001}]{Merritt2001}
{Merritt} D.,  {Ferrarese} L.,  2001, \mn@doi [\mnras] {10.1046/j.1365-8711.2001.04165.x}, \href {https://ui.adsabs.harvard.edu/abs/2001MNRAS.320L..30M} {320, L30}

\bibitem[\protect\citeauthoryear{Nowak, Saglia, Thomas, Bender, Davies  \& Gebhardt}{Nowak et~al.}{2008}]{Nowak_2008}
Nowak N.,  Saglia R.~P.,  Thomas J.,  Bender R.,  Davies R.~I.,   Gebhardt K.,  2008, \mn@doi [Monthly Notices of the Royal Astronomical Society] {10.1111/j.1365-2966.2008.13960.x}, 391, 1629–1649

\bibitem[\protect\citeauthoryear{{Pellegrini}}{{Pellegrini}}{2012}]{Pellegrini12}
{Pellegrini} S.,  2012, in {Kim} D.-W.,  {Pellegrini} S.,  eds,  Astrophysics and Space Science Library Vol. 378, Astrophysics and Space Science Library. p.~21 (\mn@eprint {arXiv} {1112.2140}), \mn@doi{10.1007/978-1-4614-0580-1_2}

\bibitem[\protect\citeauthoryear{{Peterson}}{{Peterson}}{2003}]{Peterson2003}
{Peterson} J.~R.,  2003, PhD thesis, Columbia University, New York

\bibitem[\protect\citeauthoryear{{Phillips}, {Jenkins}, {Dopita}, {Sadler}  \& {Binette}}{{Phillips} et~al.}{1986}]{Phillips1986}
{Phillips} M.~M.,  {Jenkins} C.~R.,  {Dopita} M.~A.,  {Sadler} E.~M.,   {Binette} L.,  1986, \mn@doi [\aj] {10.1086/114083}, \href {https://ui.adsabs.harvard.edu/abs/1986AJ.....91.1062P} {91, 1062}

\bibitem[\protect\citeauthoryear{Pinto et~al.,}{Pinto et~al.}{2014}]{Pinto_2014}
Pinto C.,  et~al., 2014, \mn@doi [Astronomy & Astrophysics] {10.1051/0004-6361/201425270}, 572, L8

\bibitem[\protect\citeauthoryear{{Pinto} et~al.,}{{Pinto} et~al.}{2016}]{PintoEtAl2016}
{Pinto} C.,  et~al., 2016, \mn@doi [MNRAS] {10.1093/mnras/stw1444}, \href {https://ui.adsabs.harvard.edu/abs/2016MNRAS.461.2077P} {461, 2077}

\bibitem[\protect\citeauthoryear{{Rangarajan}, {Fabian}, {Forman}  \& {Jones}}{{Rangarajan} et~al.}{1995}]{Rangarajan1995}
{Rangarajan} F.~V.~N.,  {Fabian} A.~C.,  {Forman} W.~R.,   {Jones} C.,  1995, \mn@doi [\mnras] {10.1093/mnras/272.3.665}, \href {https://ui.adsabs.harvard.edu/abs/1995MNRAS.272..665R} {272, 665}

\bibitem[\protect\citeauthoryear{{Ricci}, {Steiner}  \& {Menezes}}{{Ricci} et~al.}{2015}]{Ricci2015}
{Ricci} T.~V.,  {Steiner} J.~E.,   {Menezes} R.~B.,  2015, \mn@doi [\mnras] {10.1093/mnras/stv1156}, \href {https://ui.adsabs.harvard.edu/abs/2015MNRAS.451.3728R} {451, 3728}

\bibitem[\protect\citeauthoryear{{Sanders}}{{Sanders}}{2023}]{Sanders2023}
{Sanders} J.~S.,  2023, \mn@doi [arXiv e-prints] {10.48550/arXiv.2301.12791}, \href {https://ui.adsabs.harvard.edu/abs/2023arXiv230112791S} {p. arXiv:2301.12791}

\bibitem[\protect\citeauthoryear{{Sanders} \& {Fabian}}{{Sanders} \& {Fabian}}{2011}]{SandersFabian2011}
{Sanders} J.~S.,  {Fabian} A.~C.,  2011, \mn@doi [\mnras] {10.1111/j.1745-3933.2010.01000.x}, \href {https://ui.adsabs.harvard.edu/abs/2011MNRAS.412L..35S} {412, L35}

\bibitem[\protect\citeauthoryear{{Sarazin} \& {White}}{{Sarazin} \& {White}}{1987}]{1987SarazinWhite}
{Sarazin} C.~L.,  {White} Raymond~E. I.,  1987, \mn@doi [\apj] {10.1086/165522}, \href {https://ui.adsabs.harvard.edu/abs/1987ApJ...320...32S} {320, 32}

\bibitem[\protect\citeauthoryear{{Schweizer}}{{Schweizer}}{1980}]{1980}
{Schweizer} F.,  1980, \mn@doi [APJ] {10.1086/157870}, \href {https://ui.adsabs.harvard.edu/abs/1980ApJ...237..303S} {237, 303}

\bibitem[\protect\citeauthoryear{Shen \& Gebhardt}{Shen \& Gebhardt}{2010}]{Shen_2010}
Shen J.,  Gebhardt K.,  2010, \mn@doi [The Astrophysical Journal] {10.1088/0004-637x/711/1/484}, 711, 484–494

\bibitem[\protect\citeauthoryear{{Shurkin}, {Dunn}, {Gentile}, {Taylor}  \& {Allen}}{{Shurkin} et~al.}{2008}]{Shurkin2008}
{Shurkin} K.,  {Dunn} R.~J.~H.,  {Gentile} G.,  {Taylor} G.~B.,   {Allen} S.~W.,  2008, \mn@doi [\mnras] {10.1111/j.1365-2966.2007.12651.x}, \href {https://ui.adsabs.harvard.edu/abs/2008MNRAS.383..923S} {383, 923}

\bibitem[\protect\citeauthoryear{{Smith}}{{Smith}}{2020}]{Smith2020}
{Smith} R.~J.,  2020, \mn@doi [\araa] {10.1146/annurev-astro-032620-020217}, \href {https://ui.adsabs.harvard.edu/abs/2020ARA&A..58..577S} {58, 577}

\bibitem[\protect\citeauthoryear{{Su} et~al.,}{{Su} et~al.}{2017}]{Su2017}
{Su} Y.,  et~al., 2017, \mn@doi [\apj] {10.3847/1538-4357/834/1/74}, \href {https://ui.adsabs.harvard.edu/abs/2017ApJ...834...74S} {834, 74}

\bibitem[\protect\citeauthoryear{{Tal}, {van Dokkum}, {Nelan}  \& {Bezanson}}{{Tal} et~al.}{2009}]{Tal2009}
{Tal} T.,  {van Dokkum} P.~G.,  {Nelan} J.,   {Bezanson} R.,  2009, \mn@doi [\aj] {10.1088/0004-6256/138/5/1417}, \href {https://ui.adsabs.harvard.edu/abs/2009AJ....138.1417T} {138, 1417}

\bibitem[\protect\citeauthoryear{{Temi}, {Brighenti}  \& {Mathews}}{{Temi} et~al.}{2007}]{Temi2007}
{Temi} P.,  {Brighenti} F.,   {Mathews} W.~G.,  2007, \mn@doi [\apj] {10.1086/513690}, \href {https://ui.adsabs.harvard.edu/abs/2007ApJ...660.1215T} {660, 1215}

\bibitem[\protect\citeauthoryear{{Temi}, {Amblard}, {Gitti}, {Brighenti}, {Gaspari}, {Mathews}  \& {David}}{{Temi} et~al.}{2018}]{Temi2018}
{Temi} P.,  {Amblard} A.,  {Gitti} M.,  {Brighenti} F.,  {Gaspari} M.,  {Mathews} W.~G.,   {David} L.,  2018, \mn@doi [\apj] {10.3847/1538-4357/aab9b0}, \href {https://ui.adsabs.harvard.edu/abs/2018ApJ...858...17T} {858, 17}

\bibitem[\protect\citeauthoryear{{Temi} et~al.,}{{Temi} et~al.}{2022}]{2022Temi}
{Temi} P.,  et~al., 2022, \mn@doi [\apj] {10.3847/1538-4357/ac5036}, \href {https://ui.adsabs.harvard.edu/abs/2022ApJ...928..150T} {928, 150}

\bibitem[\protect\citeauthoryear{{Terrazas}, {Bell}, {Woo}  \& {Henriques}}{{Terrazas} et~al.}{2017}]{Terrazas2017}
{Terrazas} B.~A.,  {Bell} E.~F.,  {Woo} J.,   {Henriques} B. M.~B.,  2017, \mn@doi [\apj] {10.3847/1538-4357/aa7d07}, \href {https://ui.adsabs.harvard.edu/abs/2017ApJ...844..170T} {844, 170}

\bibitem[\protect\citeauthoryear{{Thomas}, {Fabian}, {Arnaud}, {Forman}  \& {Jones}}{{Thomas} et~al.}{1986}]{1986Thomas}
{Thomas} P.~A.,  {Fabian} A.~C.,  {Arnaud} K.~A.,  {Forman} W.,   {Jones} C.,  1986, \mn@doi [\mnras] {10.1093/mnras/222.4.655}, \href {https://ui.adsabs.harvard.edu/abs/1986MNRAS.222..655T} {222, 655}

\bibitem[\protect\citeauthoryear{{Tingay} \& {Edwards}}{{Tingay} \& {Edwards}}{2015}]{Tingay2015}
{Tingay} S.~J.,  {Edwards} P.~G.,  2015, \mn@doi [\mnras] {10.1093/mnras/stu2756}, \href {https://ui.adsabs.harvard.edu/abs/2015MNRAS.448..252T} {448, 252}

\bibitem[\protect\citeauthoryear{{Verdoes Kleijn}, {van der Marel}, {Carollo}  \& {de Zeeuw}}{{Verdoes Kleijn} et~al.}{2000}]{Verdoes2000}
{Verdoes Kleijn} G.~A.,  {van der Marel} R.~P.,  {Carollo} C.~M.,   {de Zeeuw} P.~T.,  2000, \mn@doi [\aj] {10.1086/301524}, \href {https://ui.adsabs.harvard.edu/abs/2000AJ....120.1221V} {120, 1221}

\bibitem[\protect\citeauthoryear{{White}, {Fabian}, {Johnstone}, {Mushotzky}  \& {Arnaud}}{{White} et~al.}{1991}]{WhiteEtAl1991}
{White} D.~A.,  {Fabian} A.~C.,  {Johnstone} R.~M.,  {Mushotzky} R.~F.,   {Arnaud} K.~A.,  1991, \mn@doi [\mnras] {10.1093/mnras/252.1.72}, \href {https://ui.adsabs.harvard.edu/abs/1991MNRAS.252...72W} {252, 72}

\bibitem[\protect\citeauthoryear{{Wilms}, {Allen}  \& {McCray}}{{Wilms} et~al.}{2000}]{tbabsref}
{Wilms} J.,  {Allen} A.,   {McCray} R.,  2000, \mn@doi [\apj] {10.1086/317016}, \href {https://ui.adsabs.harvard.edu/abs/2000ApJ...542..914W} {542, 914}

\bibitem[\protect\citeauthoryear{{Zhu}, {Ho}  \& {Gao}}{{Zhu} et~al.}{2021}]{ZhuHoGao21}
{Zhu} P.,  {Ho} L.~C.,   {Gao} H.,  2021, \mn@doi [\apj] {10.3847/1538-4357/abcaa1}, \href {https://ui.adsabs.harvard.edu/abs/2021ApJ...907....6Z} {907, 6}

\bibitem[\protect\citeauthoryear{{de Grijs} \& {Robertson}}{{de Grijs} \& {Robertson}}{2006}]{deGrijs2006}
{de Grijs} R.,  {Robertson} A.~R.~I.,  2006, \mn@doi [\aap] {10.1051/0004-6361:20065984}, \href {https://ui.adsabs.harvard.edu/abs/2006A&A...460..493D} {460, 493}

\bibitem[\protect\citeauthoryear{{den Herder} et~al.,}{{den Herder} et~al.}{2001}]{denHerder2001}
{den Herder} J.~W.,  et~al., 2001, \mn@doi [\aap] {10.1051/0004-6361:20000058}, \href {https://ui.adsabs.harvard.edu/abs/2001A&A...365L...7D} {365, L7}

\bibitem[\protect\citeauthoryear{{van Dokkum} \& {Conroy}}{{van Dokkum} \& {Conroy}}{2010}]{vanDokkum2010}
{van Dokkum} P.~G.,  {Conroy} C.,  2010, \mn@doi [\nat] {10.1038/nature09578}, \href {https://ui.adsabs.harvard.edu/abs/2010Natur.468..940V} {468, 940}

\bibitem[\protect\citeauthoryear{{van Dokkum} \& {Conroy}}{{van Dokkum} \& {Conroy}}{2024}]{2024vandokkumconroy}
{van Dokkum} P.,  {Conroy} C.,  2024, \mn@doi [arXiv e-prints] {10.48550/arXiv.2407.06281}, \href {https://ui.adsabs.harvard.edu/abs/2024arXiv240706281V} {p. arXiv:2407.06281}

\makeatother
\end{thebibliography}

%\begin{thebibliography}{99}
%\end{thebibliography}

%%%%%%%%%%%%%%%%%%%%%%%%%%%%%%%%%%%%%%%%%%%%%%%%%%

%%%%%%%%%%%%%%%%% APPENDICES %%%%%%%%%%%%%%%%%%%%%

\appendix

\section{Alternative Models Discussed in Spectral Fitting}\label{AppendixA}

\subsection{Single Temperature Model}
The Single-Temperature model describes a gas at a single temperature, without a cooling flow. Only the components representing Galactic absorption, Gaussian smoothing, and a single-temperature gas are included in the model. In XSPEC, it is defined as \textsc{tbabs*(gsmooth*apec)}. 

\subsection{Two Temperature Model}
The Two-Temperature model describes a mixture of gases at two different temperatures, again without a cooling flow. In XSPEC it is defined similarly to the Single-Temperature model, but with two \textsc{apec} components instead of one: \textsc{tbabs*(gsmooth*apec+gsmooth*apec)}. Each \textsc{apec} component has the same abundance, but its own Gaussian smoothing component; the hotter temperature component is assumed to be more widespread and consequently requires more smoothing.
{\renewcommand{\arraystretch}{1.4}
\begin{table*}
\normalsize
\centering
 \begin{tblr}{c c c c c c c c c c}
 \hline
 Target & $N_\text{H}'$ & $kT$ & $Z$ & $Norm$ & $CFrac$ & $N_\text{H}$& $\dot{M}$ & $\dot{M}_\text{u}$\\ 
 \hline
  & $10^{22}$ cm\textsuperscript{-2}& keV & $Z_{\odot}$& $10^{-4}$ &  & $10^{22}$ cm\textsuperscript{-2}& $M_\odot$ yr\textsuperscript{-1} &  $M_\odot$ yr\textsuperscript{-1}\\
 \hline
 NGC 1316 & 0.0199 & $1.00^{+0.05}_{-0.12}$ & $0.21^{+0.03}_{-0.04}$ & 7.8e-3 & $0.90^{+0.06}_{-0.01}$ & 3.8 & $4.4^{+3.1}_{-2.2}$ & $0.52 \pm 0.04$\\
 NGC 1332 & 0.0214 & $0.67^{+0.04}_{-0.05}$ & $0.11^{+0.03}_{-0.02}$ & 2.4 & $1_{-0.04}$ & 4.7 & $7.9^{+2.8}_{-0.4}$ & $0.27^{+0.06}_{-0.05}$\\
 NGC 1404 & 0.0138 & $0.69 \pm 0.01$ & $0.21\pm 0.01$ & 20 & $0.97^{+0.02}_{-0.03}$ & 2.2 & $8.0\pm 0.3$ & $0.62\pm 0.14$\\
 NGC 4552 & 0.0266 & $0.70\pm0.02$ & $0.072^{+0.007}_{-0.006}$ & 7.3 & $0.97^{+0.03}_{-0.15}$ &4.6 & $0.72^{+0.6}_{-0.1}$ & $0.036^{+0.004}_{-0.006}$\\
 NGC 4636 & 0.0182 & $0.76\pm0.02$ & $0.26\pm0.02$ & 33 & $1_{-0.24}$ & 0.33 & $0.91^{0.23}_{-0.22}$ & $0.32^{+0.06}_{-0.05}$\\
 NGC 4649 & 0.0202 & $0.91\pm0.01$ & $0.32\pm0.02$ & 23 & $1_{-0.01}$ & 9.9 & $4.0^{+1.5}_{-1.9}$ & negligible\\
 IC 1459 & 0.00967 & $0.73^{+0.16}_{0.07}$ & $0.09^{+0.03}_{-0.02}$ & 1.6 & $0.91^{+0.06}_{-0.39}$ & 6.2 & $3.2^{+2.9}_{-2.5}$ & $0.33^{+0.05}_{-0.06}$\\
 \hline
\end{tblr}
\caption{Model parameters for the \textit{cstat} analysis. The redshifts for each galaxy are as given in \hyperref[Table2]{Table 2}. $kT$ is the temperature of the \textsc{apec} model component; $Z$ is the abundance relative to solar abundance. Where $Cfrac$ was fitted as 1, there is no upper error, so only a lower bounds on the error is given. The errors represent the 90\% confidence intervals using $\chi^2$.}
\label{TableB}
\end{table*}
}

\section{\textit{Cstat} Refits}\label{AppendixB}

The same approach to fitting as described in \hyperref[SA]{Section 2} was adopted, but this time with \textit{cstat} as the fit statistic instead of $\chi^2$. The parameters determined in this case agree with those found with the $\chi^2$ analysis within their quoted errors, indicating that the model is a good fit.

There was no noticeable improvement to the quality of the fit when fitting with \textit{cstat} compared to $\chi^2$; in fact, the contour plots in the \textit{cstat} analysis were significantly more granular. This gives confidence in the goodness of the $\chi^2$ fit.

%%%%%%%%%%%%%%%%%%%%%%%%%%%%%%%%%%%%%%%%%%%%%%%%%%

% Don't change these lines
\bsp	% typesetting comment
\label{lastpage}
\end{document}